\documentclass[12pt]{article}
\usepackage{amssymb,amsmath}
\usepackage{amsfonts,amsthm,fancybox,amstext,amscd,array,bbold,bm,bbm,slashed,graphicx,mathrsfs,subfigure}
\usepackage{mathrsfs}
\usepackage{color}

\makeatletter

\@addtoreset{equation}{section}
\def\section{\@startsection {section}{1}{\z@}{-3.ex plus -1ex minus
 -.2ex}{1.0ex plus .2ex}{\large\bf}}
\def\subsection{\@startsection{subsection}{2}{\z@}{-2.45ex plus%
 -1ex minus -.2ex}{0.5ex plus .2ex}{\bf}}

\def\cd{\!\cdot\!}

\newcommand{\NN}{\mathbb{N}}

\newcommand{\RR}{\mathbb{R}}

\newcommand{\HH}{\mathbb{H}}

\def\bmz{\left(\begin{array}{2,2}}
\def\emz{\end{array}\right)}
\def\bmd{\left(\begin{array}{3,3}}
\def\emd{\end{array}\right)}

\newcommand{\ep}{\epsilon} 
\newcommand{\en}{\omega}    
\newcommand{\myA}{A^{s}_{i}}
\newcommand{\myAj}{A^{s}_{j}}
\newcommand{\myphi}{\phi^{s}}
\newcommand{\myF}{F^{s}_{ij}}
\newcommand{\myD}{D^{s}}

\DeclareMathAlphabet{\mathpzc}{OT1}{pzc}{m}{it} 
\newcommand{\myW}{\mathpzc{Q}}

\newcommand{\myv}{\mathpzc{p}} 
\newcommand{\myvv}{\mathpzc{q}} 
\newcommand{\x}{\hat{\mathbf{x}}}
\newcommand{\bx}{\mathbf{x}}
\newcommand{\by}{\mathbf{y}}

\newcommand{\Lv}{\mathbf{L}}
\newcommand{\tv}{\mathbf{t}}
\newcommand{\ev}{\mathbf{e}}
\newcommand{\sv}{\mathbf{s}}
\newcommand{\basis}{b}

\newcommand{\nts}[1]{}
\newcommand{\nag}[1]{\emph{#1}}

\begin{document}
\parskip 5pt
\parindent 0pt
\begin{flushright}
EMPG-10-20\\
\end{flushright}

\begin{center}
\baselineskip 24 pt
{\Large \bf    On resonances and bound states of the \\ 't Hooft-Polyakov monopole} 

\vspace{1cm}
{ K.~M.~Russell\footnote{\tt kmr35@cantab.net}
and    B.~J.~Schroers\footnote{\tt bernd@ma.hw.ac.uk} \\
Department of Mathematics and Maxwell Institute for Mathematical Sciences, \\Heriot-Watt University, 
Edinburgh EH14 4AS, United Kingdom } \\

\vspace{0.5cm}

{December  2010}

\end{center}

\begin{abstract}
\baselineskip 12pt
\noindent
We  present a systematic approach to the linearised Yang-Mills-Higgs equations in the background 
of a 't Hooft-Polyakov monopole and use it to unify and extend  previous studies  of their spectral properties.  We show that 
a quaternionic formulation allows for a compact and efficient treatment of the linearised equations in the BPS limit of vanishing Higgs self-coupling,
and use it to study both scattering and bound states. We focus on the sector of vanishing generalised angular momentum
and  analyse it  numerically,  putting zero-energy bound states, Coulomb bound states and infinitely many 
Feshbach resonances into a  coherent picture. We also consider the linearised Yang-Mills-Higgs equations with   non-vanishing Higgs self-coupling and  confirm the occurrence of  Feshbach resonances in this situation. 

\end{abstract}
\section{Introduction}

Solitons - spatially localised and stable solutions of non-linear differential equations - are  widely used in modelling physical and biological phenomena. In many of these applications, it is also of interest to study the behaviour of small perturbations around the soliton. In the context of quantum field theory,
the small perturbations are interpreted as particles after quantisation \cite{Rajaraman}. In that case,  studying perturbations around a soliton amounts to studying the interaction of these (quantum) particles with the classical soliton. In Skyrme's model for  nuclei \cite{Skyrme1, Skyrme2, ManSut:04}, for example, the solitons describe nuclei and  the small  perturbations correspond, after quantisation, to pions.   Their properties in the background of  a  Skyrmion  can  therefore be interpreted in terms  of  pion-nucleus scattering.

Mathematically, studying perturbations of a  soliton amounts to studying the spectral properties of the linear operator obtained by linearising the soliton equations around the soliton.  It turns out that these spectral properties are often interesting, see \cite{Rajaraman} for a textbook treatment of some examples.  A particularly interesting example is provided by  magnetic monopoles in $SU(2)$ Yang-Mills-Higgs (YMH) theory.  It has long been known that  the linearised YMH equations in the background of a monopole have interesting zero-modes in the BPS limit \cite{JacReb:75,Weinberg,Callias} and that they support infinitely many bound states \cite{BaiTro:81}. Scattering of fermions off the monopole has also been studied extensively \cite{Rubakov1,Rubakov2,Callan,MarMuz}. 
However, fairly recently    interesting  new scattering phenomena were observed. In \cite{FodRac:03} and, more recently and in more detail in \cite{FodRac:08},  Fodor and R\'acz studied spherically symmetric but non-linear perturbations of the 't Hooft-Polyakov monopole, in the BPS limit.  The authors studied the evolution of such  excitations numerically   and found that  the monopole  holds on to a significant fraction of the energy from
the initial excitation for a surprisingly long time. In particular the amplitude of the excitation decays
as $t^{-5/6}$ for late times.

Motivated by this work, Forg\'acs and Volkov carried out  a
perturbative analysis \cite{ForVol:03} of  the 't Hooft-Polyakov monopole,  still  in the BPS limit and preserving the hedgehog form of the monopole.  Their
 linearised equations of motion are  a   system of  two weakly coupled, second 
 order ordinary differential equations. 
By using the dispersion relation, one of these channels is seen to be massive and one massless.
 Some intuition about  the system is gained by artificially decoupling it and considering only the massive channel. The potential appearing  in it 
 has an (attractive) Coulomb tail, so that it possesses
 infinitely many bound states approaching a critical value, beyond
 which there is a continuous spectrum. On re-coupling and considering the full system, a
 phase shift analysis of the massless channel shows that the infinity
 of bound states in the massive channel  turn into  an infinity of resonances in the coupled system. The  energy held by the resonances leaks out slowly to the massless channel, leading to  the  long-lived excitation  observed  by Fodor and R\'acz.

The aim of this work is to extend the results of Forg\'acs and Volkov and to place it in the context of other spectral properties of the linearised YMH equations. We develop a quaternionic formalism for studying the linearised YMH equation in the case of vanishing Higgs self-coupling (i.e. in the BPS limit). We show that this quaternionic formalism  allows for a systematic treatment of perturbations, organised in terms of the eigenvalues of the generalised angular momentum operator (combining orbital angular momentum, spin and isospin).  For vanishing generalised angular momentum we  recover the  equations studied by  Forg\'acs and Volkov but also  find another system, consisting of  two coupled, second 
 order ordinary differential equations. These have bound states, already found  by Bais and Troost in \cite{BaiTro:81}.  The  two systems found for  vanishing generalised angular momentum thus already display a wealth of interesting spectral properties, including bound states at zero energy, bound states embedded in the continuum and resonance scattering. Moreover, the BPS condition allows one to map either 
of the systems into equivalent but sometimes simpler systems using a first order differential operator (essentially a supersymmetry charge, but we do not consider the fully supersymmetric theory here). This turns out to explain some of the surprising features we find.  

We also consider the case of non-vanishing Higgs self-coupling $\lambda$.  In this regime,
our quaternionic formalism is no longer effective. Thus we do not study general perturbations but instead focus on  the generalisation of the system studied by Forg\'acs and Volkov when  $\lambda\neq 0$. We find that the Coulomb tail in the massive channel is replaced by an attractive  $1/r^2$ potential. This is strong enough to support infinitely many bound states after decoupling (by hand), and our numerical analysis suggests that it will also produce infinitely many resonances  in the coupled systems. We conclude that the 
qualitative features found by Forg\'acs and Volkov survive the ``switching on'' of $\lambda$.


The paper is organised as follows. In Sect.~2,  we study the general form of the linearised YMH equation around a background which 
satisfies the first order BPS equation (which  includes the 't Hooft-Polyakov monopole in the BPS limit). We introduce a quaternionic language
and show that the linearised YMH equations for stationary time-dependent perturbations can be expressed as a quaternionic wave equation, supplemented by a background gauge condition.  For non-zero kinetic energy there is a second, equivalent form  of this wave equation, related to the original one by the application of a Dirac-type operator. Our strategy for studying  the  linearised YMH equations 
is therefore to study  the quaternionic wave equation and to check if solutions satisfy the background gauge condition. 

In Sect.~3, we   carry out a partial wave analysis of the quaternionic wave equation and  derive   two systems of two second order ordinary differential equations 
which arise in the sector with vanishing generalised angular momentum. We also derive the form of the alternative but equivalent systems obtained by acting with
 the Dirac-type operator of the previous section.  Sect.~4  contains a detailed, numerical investigation of the two systems found in Sect.~3.  Even though these systems  
 look superficially very similar, their spectral properties are quite different.   One of the systems 
  is the hedgehog system already discussed in  \cite{ForVol:03}; we briefly repeat the analysis of this system, using it to set out our conventions.  We then observe that the other system  decouples, after application of the Dirac-type operator, into one channel which supports bound states and another which only has scattering states.   We are  able to relate the bound states to those discussed in \cite{BaiTro:81}. The scattering states do not satisfy the background gauge condition and therefore are not valid bosonic states in the linearised theory,  but  we point out their relation to the fermionic scattering states studied  in \cite{MarMuz}. In  Sect.~5, we continue our study of $SU(2)$ monopoles, but allow for
non-zero Higgs self-coupling. We perturb around the background of the non-BPS  't Hooft-Polyakov monopole and find scattering
resonances in the linearised hedgehog fields.  Finally,  in Sect.~6  we discuss possible extensions of our work and the interpretation of our results in the context of electric-magnetic duality.

\section{Perturbing the BPS Monopole}
\subsection{The BPS Monopole}

Much  background material  for this section can be found in  the textbook  \cite{ManSut:04}, to which we refer for details. We work  on four-dimensional  Minkowski space-time with  coordinates $x^{\mu}$, $\mu = 0,1,2,3$ and  Minkowski metric $\eta_{\mu\nu} =\text{diag}(1,-1,-1,-1)$. We denote the time coordinate by $x^0$ or  $t$ and write three-component Euclidean vectors with bold letters,  e.g.
$\bx=(x_1,x_2,x_3)$.  We write the inner product as $\bx\cdot \by$ and  $r$ for the spatial radial coordinate, i.e. $r=|\bx| = \sqrt{x_1^2+x_2^2 +x_3^2}$. 
The YMH model we are interested in  has gauge
group $SU(2)$, often referred to as isospin symmetry. Monopoles emerge
when the Higgs mechanism breaks the $SU(2)$ symmetry to $U(1)$.  In the notation and nomenclature of this paper we treat this $U(1)$  as the gauge group of Maxwell electrodynamics.  We work in units where the speed of light and the gauge coupling are 1. 

The fields of the YMH model are an
$SU(2)$ gauge potential $A_{\mu}$, coupled to a
Higgs field $\phi$. Both take values in the Lie algebra $\mathfrak{su(2)}$ and   transform in  the adjoint representation of
$SU(2)$.  For  the Lie algebra $\mathfrak{su(2)}$, we use the basis ${t_a= -\frac{i}{2}\tau_a}$, where $\tau_a, a=1,2,3$,  are the
Pauli matrices, with brackets  $[t_a,t_b]=\ep_{abc}t_c$, noting that this is \emph{not} the
 same convention as  \cite{ManSut:04}.  We will also need an inner product $\langle, \rangle$ on $\mathfrak{su(2)}$, which we normalise so that $\langle t_a, t_b\rangle =\delta_{ab}$. The
covariant derivative is $D_{\mu} = \partial_{\mu} + [A_{\mu},~]$
and the Yang-Mills field strength tensor, or curvature 2-form, is $F_{\mu\nu} = \partial_{\mu}A_{\nu} -
\partial_{\nu}A_{\mu}+ [A_{\mu},A_{\nu}]$. From this we can extract the non-abelian
electric field $E_i=F_{0i}$ and the non-abelian magnetic field
\begin{equation}
\label{magdef}
B_i=-\frac{1}{2}\ep_{ijk}F_{jk}, \quad i,j,k=1,2,3.
\end{equation}

The YMH Lagrangian density is 
\begin{equation}\label{Lag}
    \mathcal{L} =
                   -\frac{1}{4}\langle F_{\mu\nu}F^{\mu\nu}\rangle
                       + \frac{1}{2}\langle D_{\mu}\phi D^{\mu}\phi\rangle
                        - \frac{\lambda}{4}(1-|\phi|^2)^2,
\end{equation}
where $|\phi|^2=\langle \phi, \phi \rangle$. The equations of
motion derived from the Lagrangian density are 
\begin{subequations}\label{EoM}
\begin{align}
D_{\mu}D^{\mu}\phi &= \lambda(1-|\phi|^2)\phi, \label{EoM1}\\
D_{\mu}F^{\mu\nu} &= [D^{\nu}\phi,\phi].\label{EoM2}
\end{align}
\end{subequations}

Static configurations play an important role in this paper as background configurations. 
For such configurations we work in the temporal gauge
$A_0=0$, and assume time-independence of the remaining fields $A_i$ and $\phi$. Sometimes we collect
 the  static gauge field into a spatial one-form $A=A_idx^i$,  and write $(A,\phi)$ for
 the field configuration.  For such configurations, the energy computed from the Lagrangian \eqref{Lag} can only 
be finite if we impose the boundary condition
\begin{align}
\label{inftycond}
 |\phi|\rightarrow 1 \quad \text{as} \quad r \rightarrow \infty.
\end{align}
The equations of motion
(\ref{EoM}) then reduce to 
\begin{subequations}\label{SEoMH}
\begin{align}
D_{i}D_{i}\phi &=- \lambda(1-|\phi|^2)\phi, \label{SEoM1H}\\
D_{i}F_{ij} &= -[D_j\phi,\phi].\label{SEoM2H}
\end{align}
\end{subequations}
The 't Hooft-Polyakov static monopole solution found in  \cite{tHo:74, Pol:74}
has the hedgehog form
\begin{equation}
A_i(\bx)=\frac{x_k}{r^2}(1-W(r))\ep_{aik}t_a,  \quad \phi(\bx) = \frac{H(r)}{r^2}x_at_a. \label{hedgehog}
\end{equation}
 Regularity at the origin requires that $W(0)=1$ and $H(0)=0$, while  the boundary conditions at infinity \eqref{inftycond} are satisfied if we require  $H\to -r$ as $r\to \infty$. To recover a more standard  definition of the Higgs field $H(r)$ we can scale by $-r$. Our definition of $H(r)$ follows the conventions of Forg\'acs and Volkov  in  \cite{ForVol:03} for the hedgehog ansatz, for better comparison with their work. 

$H(r)$ and $W(r)$ satisfy differential equations derived from (\ref{SEoMH}),
\begin{subequations}\label{2ndOrderWHeqnsH}
\begin{align} 
\left(-r^2\frac{d^2}{dr^2}+W^2+H^2-1\right)W&=0, \\
\left(-r^2\frac{d^2}{dr^2}+2W^2-\lambda(r^2-H^2)\right)H&=0.
\end{align}
\end{subequations}

The BPS limit amounts to setting  $\lambda= 0$ in (\ref{Lag}) but maintaining the boundary condition on $\phi$. As the mass of the Higgs is proportional to $\sqrt{\lambda}$,  the Higgs field is massless in this  limit. The static field equations (\ref{SEoMH}) become
\begin{subequations}\label{SEoM}
\begin{align}
D_{i}D_{i}\phi &=0, \label{SEoM1}\\
D_{i}F_{ij} &= -[D_j\phi,\phi],\label{SEoM2}
\end{align}
\end{subequations}
while the equations (\ref{2ndOrderWHeqnsH}) reduce to 
\begin{subequations}\label{2ndOrderWHeqns}
\begin{align}
\left(-r^2\frac{d^2}{dr^2}+W^2+H^2-1\right)W&=0, \\
\left(-r^2\frac{d^2}{dr^2}+2W^2\right)H&=0.
\end{align}
\end{subequations}
These equations have an analytic solution found by Prasad and Sommerfield in  \cite{PraSom:75},
\begin{equation}\label{WH}
H(r)=1-r\coth(r), \quad W(r)=\frac{r}{\sinh(r)}.
\end{equation}
The corresponding field configuration is called the Bogomol'nyi-Prasad-Sommerfield (BPS) monopole.

Further insight into the  BPS limit can be gained from considering  the static energy 
\begin{equation}\label{staticenergy}
E = \frac{1}{2}\int \langle B_i, B_i\rangle  + \langle D_i\phi,  D_i\phi\rangle d^3x,
\end{equation}
and rearranging it to 
\begin{equation}\label{staticenergy2}
E = \frac{1}{2}\int \langle B_i+D_i\phi, B_i+D_i\phi \rangle d^3x - \int \partial_i\langle B_i, \phi\rangle  d^3x,
\end{equation}
where we used  the Bianchi identity $D_iB_i=0$. Using Stokes' law and the quantisation of magnetic flux, one finds
\begin{equation}\label{staticenergy3}
E = \frac{1}{2}\int \langle B_i+D_i\phi, B_i+D_i\phi\rangle d^3x + 2\pi N,
\end{equation}
where $N$ is an integer called the monopole number.
Then for $N > 0$ (i.e. monopoles as opposed to anti-monopoles), the energy is bounded by
\begin{equation}\label{energybound}
E \geq 2\pi N.
\end{equation}
This bound is saturated  when the  Bogomol'nyi equation  \cite{Bog:76}
\begin{equation}
\label{BE}
B_i+D_i\phi=0
\end{equation}
holds.  
The exact monopole solution (\ref{WH}) satisfies  this equation, 
and $W(r)$ and $H(r)$ thus  satisfy first order equations derived from (\ref{BE}), which we will use later:
\begin{subequations}\label{WHeqns}
\begin{align}
rH'(r)&=W(r)^2+H(r)-1,\\
rW'(r)&=W(r)H(r).
\end{align}
\end{subequations}
It is straightforward to check that the first order partial differential equations \eqref{BE} imply the second order equations \eqref{SEoM}, and  that the first order ordinary differential equations \eqref{WHeqns} imply the second order equations  \eqref{2ndOrderWHeqns}. 

\subsection{Static linearisation}
We fix a static background configuration $(A^s,\phi^s)$, assumed to satisfy  the Bogomol'nyi equation \eqref{BE},  write $\myD_i$ for the  covariant derivative $\myD_i = \partial_i + [\myA,~]$ and likewise  $\myF$ for the curvature of $A^s_i$.  Although we are ultimately interested in time-dependent perturbations, we  begin by  considering  static  perturbations  $(a,\varphi)$ of  $(A^s,\phi^s)$
and insert
\begin{align}
\label{perturbgaugefields}
A_i  = \myA+a_i, 
\qquad 
\phi = \myphi+\varphi, 
\end{align}
into \eqref{SEoM}. Substituting into (\ref{SEoM1}),  applying the static background gauge condition and collecting linear terms in the perturbation we find
\begin{equation}
D_{i}D_{i}\phi  \simeq \myD_{i}\myD_{i}\varphi + [a_i,\myD_i\myphi] + \myD_i[a_i,\myphi] = 0. \\
\end{equation}
To linearise (\ref{SEoM2}), we need to use that, to linear order,
\begin{equation}
F_{ij}\simeq \myF + \partial_i a_j - \partial_j a_i + [a_i,\myAj] + [\myA,a_j].
\end{equation}
Then substituting into (\ref{SEoM2}), the left hand
side is
\begin{equation}
\label{LHS}
D_iF_{ij} \simeq \myD_i\myF + \myD_i\myD_ia_j - \myD_i\myD_ja_i + [a_i,\myF].
\end{equation}
Likewise the right hand side of (\ref{SEoM2}) becomes 
\begin{equation}\label{RHS}
-[D_j\phi,\phi] \simeq [\myphi,\myD_j\myphi] + [\myphi,\myD_j\varphi] + [\varphi,\myD_j\myphi] +
[\myphi,[a_j,\myphi]].
\end{equation}
Putting (\ref{LHS}) and (\ref{RHS}) together and  applying the static
equation (\ref{SEoM2}) we obtain the linearisation
\begin{equation}
 \myD_i\myD_i a_j - \myD_i\myD_j a_i + [a_i,\myF] \simeq  [\myphi,\myD_j\varphi] + [\varphi,\myD_j\myphi] +[\myphi,[a_j,\myphi]].
\end{equation}

In most of  the remainder of the paper  we will only use the 
covariant derivative $\myD_i$  and the field strength $\myF$ associated with a fixed static background configuration.  In order to simplify notation we therefore drop the superscripts on $\myA,\myphi$,
 and also on the associated covariant derivative and curvature. In this notation, 
the linearised Yang Mills Higgs equations for static fields are 
\begin{subequations}\label{LSEoM}
\begin{align}
D_{i}D_{i}\varphi + [a_i,D_i\phi] + D_i[a_i,\phi] &= 0, \label{LSEoM1}\\
D_iD_ia_j - D_iD_ja_i + [a_i,F_{ij}] &= [\phi,D_j\varphi] + [\varphi,D_j\phi] +[\phi,[a_j,\phi]].\label{LSEoM2}
\end{align}
\end{subequations}

We can  similarly linearise the Bogomol'nyi equations. Substituting  \eqref{perturbgaugefields}
 into the Bogomol'nyi equation (\ref{BE}),  linearising, using the Bogomol'nyi equation  and re-naming 
 $\phi^s \rightarrow \phi, A^s_i\rightarrow A_i$ gives
\begin{equation}\label{LBE}
\epsilon_{ijk}D_ja_k=D_i\varphi +[a_i,\phi].
\end{equation}
There are infinitely many solutions of this equation  with $a_i=-D_i\vartheta$, $\varphi=[\vartheta,\phi]$, where $\vartheta$ is an arbitrary function on $\mathbb{R}^3$ with  values in $\mathfrak{su(2)}$. These do not physically change the original static solution, as they are  infinitesimal gauge transformations. We can exclude such solutions by requiring that perturbations $(a_i,\varphi)$ satisfy
\begin{equation}
\int\left(\langle a_i, D_i\vartheta\rangle +\langle \varphi, [\vartheta,\phi]\rangle \right)d^3x=0,
\end{equation}
for all $\vartheta$ which are non zero on  a closed and bounded subset of $\mathbb{R}^3$. The requirement of compact support means that we can integrate by parts and rearrange to obtain the background gauge condition 
\begin{equation}\label{backgd_gauge}
D_ia_i+[\phi,\varphi]=0.
\end{equation}

Interestingly, the linearised Bogomol'nyi equations  together with the background gauge condition  imply the linearised YMH equations (just as solutions of the Bogomol'nyi equation (\ref{BE}) are solutions to the static YMH equations (\ref{SEoM})).
 To see this,  we apply $D_i$ to (\ref{LBE}) and then  use \eqref{BE} to  obtain (\ref{LSEoM1}).
In order to derive (\ref{LSEoM2}), we apply $D_j$ to the background gauge condition 
(\ref{backgd_gauge}) and  $[\phi,\cdot]$ as well as $D_l$ to  the linearised Bogomol'nyi equation  (\ref{LBE}). The algebra is a little tedious, and makes repeated use of \eqref{BE}. We will give a much quicker derivation in the next section.

\subsection{Quaternionic Formulation}
\label{QuaternionicFormulation}
We will now show that the language of  quaternions is very convenient for studying the linearised equations of the previous section.  We denote the set of all quaternions as $\HH$ and introduce the usual basis
$e_{\alpha}$, $\alpha=1,2,3,4$. The real unit quaternion $e_4$ commutes with all quaternions  and is often written as a
$1$ or omitted. The remaining (imaginary) quaternions satisfy
\begin{equation}
e_i e_j = -\delta_{ij} + \epsilon_{ijk}e_k \qquad (i,j,k = 1,2,3). 
\end{equation}
We can identify $e_j = -i\sigma_j$,  where $\sigma_j$ are again the  Pauli matrices (but not denoted $\tau_i$ here in order to avoid confusion with the isospin Lie algebra), and $e_4$  with the $2\times2$
identity matrix $\mathbbm{1}_2$. The conjugates are
\begin{equation}
\bar{e}_i=-e_i, \quad \bar{e}_4=e_4.
\end{equation}

We combine the gauge fields $A_i$ and $\phi$ into a
quaternion-valued field
\begin{equation}
\label{quatback}
\myW=A_ie_i+\phi.
\end{equation} 
Since $A_i$ is an  isovector-vector and $\phi$ is an isovector-scalar, we can view this field as a map
\[
\myW: \RR^3 \rightarrow \HH \otimes \mathfrak{su}_2.
\]

Next, we define Dirac-type derivative operators
\begin{equation}
\label{Diracops}
\slashed{D}=D_ie_i+[\phi,~], \quad \slashed{D}^{\dag}=D_ie_i-[\phi,~],
\end{equation} 
which act on functions $\myvv: \RR^3\rightarrow \HH \otimes \mathfrak{su}_2$ by quaternionic multiplication on the quaternions and by commutator on the isospin part $\mathfrak{su}_2$. These operators are closely related to the Bogomol'nyi equation for the background field $\myW$. Note that 
\begin{equation}
\slashed{D}^\dagger \slashed{D} = -D_i^2 - \phi^2 + (D_i\phi - B_i)e_i, \quad \slashed{D} \slashed{D}^\dagger = -D_i^2  -\phi^2 + (-D_i\phi - B_i)e_i,
\end{equation}
so that, for BPS monopoles,
\begin{equation}
\label{BPSdirac}
\slashed{D}^\dagger \slashed{D} =- D_i^2 - \phi^2 + 2D_i e_i, \quad \slashed{D} \slashed{D} ^\dagger= -D_i^2 - \phi^2, 
\end{equation}
by virtue of \eqref{BE}.

It turns out that the linearised YMH and BPS equations can both be expressed very compactly in terms of these operators.
To see this,  let 
\begin{equation}
\label{quatexp}
 \myvv=a_ie_i+\varphi.
\end{equation} 
We look for the quaternionic expression equivalent to the linearised Bogomol'nyi
equation (\ref{LBE}) and observe that
\begin{equation}\label{QLBEexpression}
\begin{split}
\slashed{D}\bar{\myvv}&=D_ia_i + [\phi,
\varphi] + e_i(D_i\varphi - \ep_{ijk}D_ja_k - [\phi,a_i]).
\end{split}
\end{equation}
The real part of (\ref{QLBEexpression}) vanishing is precisely the background
gauge condition (\ref{backgd_gauge}), while setting the complex part to zero
is equivalent to (\ref{LBE}). 
The linearised Bogomol'nyi equation (\ref{LBE})  togther with the background gauge condition therefore have the following, very simple quaternionic
formulation 
\begin{equation}\label{QLBE}
\slashed{D}\bar{\myvv} = 0.
\end{equation}
This was used extensively  in  \cite{ManSch:92}.

We now express  the  linearised
YMH  equations (\ref{LSEoM}) in quaternionic notation, which has not been previously considered. We expect it to be some second order equation in $\slashed{D}$,
since the linearised Bogomol'nyi equation is a first order equation in $\slashed{D}$.
In fact, we will now show that the quaternionic equation
\begin{equation}\label{QLYMH}
\slashed{D}^{\dag}\slashed{D}\bar{\myvv}=0
\end{equation} 
 is equivalent to the linearised field equations \eqref{LSEoM1} and \eqref{LSEoM2}, provided the background gauge condition \eqref{backgd_gauge} holds.

In order to prove our claim  we first note a   number of useful relations.  Using the definition of the curvature as the commutator of covariant derivatives,  as well as the Bogomol'nyi equation \eqref{BE} and the definition \eqref{magdef} of the  non-abelian magnetic field,  one finds
\begin{equation}
\ep_{lim}D_lD_i\varphi = \frac{1}{2}\ep_{lim}[D_l,D_i]\varphi  
= \frac{1}{2}\ep_{lim}[F_{li},\varphi] 
= [B_m,\varphi]   
= [\varphi, D_m\phi].
\end{equation}
The Leibniz rule, again with the Bogomol'nyi equation \eqref{BE} and the definition \eqref{magdef} of the  non-abelian magnetic field, gives
\begin{equation}
\ep_{lim}D_l[\phi,a_i]= [F_{im},a_i]+ \ep_{lim}[\phi,D_la_i].
\end{equation}
Then we compute
\begin{equation}\label{QLYMHexpression}
\begin{split}
\slashed{D}^{\dag}\slashed{D}\bar{\myvv}&=
\slashed{D}^{\dag}(D_ia_i + [\phi, \varphi]) + 
\slashed{D}^{\dag} e_i(D_i\varphi - \ep_{ijk}D_ja_k - [\phi,a_i])\\
&=\slashed{D}^{\dag}(D_ia_i + [\phi, \varphi]) \\
&\quad - D_{i}D_{i}\varphi - [a_i,D_i\phi] -
D_i[a_i,\phi]\\
&\quad + D_iD_ia_je_j - D_iD_ja_ie_j + e_j[a_i,F_{ij}] -
e_j[\phi,D_j\varphi] - e_j[\varphi,D_j\phi] - e_j[\phi,[a_j,\phi]].
\end{split}
\end{equation}
The terms after the last equality sign are, in the first line, the operator $\slashed{D}^{\dag}= 
e_lD_l-\phi$ applied to the 
  the background gauge expression \eqref{backgd_gauge},
in the second line the linearised field equations (\ref{LSEoM1}),  and in the third line the linearised field equations
 (\ref{LSEoM2}).  Thus,  with the  background gauge condition $D_ia_i + [\phi, \varphi]=0$  imposed, the quaternionic equation \eqref{QLYMH} and the linearised  static YMH equations \eqref{LSEoM} are equivalent, as claimed.

In the quaternionic notation it is obvious that  the first order equation 
$\slashed{D}\bar{\myvv} = 0$ (combining background gauge condition and linearised Bogomol'nyi equation) implies the second order equation
$\slashed{D}^{\dag}\slashed{D}\bar{\myvv}=0$ (and hence the linearised static field equations, since the background gauge is in place).

\subsection{Time dependent perturbations}\label{BPSTimeDependent}
 We now introduce  time-dependent perturbations around a static configuration $(\myA, \myphi)$ 
 satisfying  \eqref{BE},  using the following  stationary ansatz
\begin{equation}
 A_i(t,\bx)=A_i^s(\bx) + a_i(\bx)e^{i\omega t}, \qquad \phi(t,\bx)=\myphi(\bx)+ \varphi(\bx)e^{i\omega t}.
\end{equation}
 We 
recall that we work  in the  temporal gauge $A_0=0$ and the BPS limit $\lambda=0$.
Inserting the ansatz in the  YMH equations (\ref{EoM}), linearising and re-naming again $\phi^s\rightarrow \phi, A^s_i\rightarrow A_i$, we find 
\begin{subequations}\label{LSEoMT}
\begin{align}
D_{i}D_{i}\varphi + [a_i,D_i\phi] + D_i[a_i,\phi] &=
-\omega^2\varphi, \label{LSEoM1T} \\
D_iD_ia_j - D_iD_ja_i + [a_i,F_{ij}] &= [\phi,D_j\varphi] + [\varphi,D_j\phi] +[\phi,[a_j,\phi]] -\omega^2 a_j. \label{LSEoM2T}
\end{align}
\end{subequations}
For $\en=0$ we recover the static linearised YMH equation (\ref{QLYMH}), as one would expect.

With the results of the previous two subsections, we  can express the linearised equations in quaternionic language. Provided the background gauge condition \eqref{backgd_gauge} holds, the following
simple equation
\begin{equation}\label{master}
\slashed{D}^{\dag}\slashed{D} \bar{\myvv}=\en^2 \bar{\myvv}
\end{equation}
is equivalent to the stationary linearised YMH equations \eqref{LSEoMT}. 
This observation is fundamental for the remainder of this paper, and the foundation of our strategy for 
investigating \eqref{LSEoMT} by studying \eqref{master} and then imposing the background gauge condition.

Like all differential equations, the equation (\ref{master}) can be written as a first order system.  In this case, it takes the form of a Dirac equation:
\begin{equation}
\bigg(
 \begin{array}{cc}
   \partial_t & \slashed{D}^{\dag} \\
\slashed{D} & -\partial_t 
\end{array} \bigg)
\bigg(\begin{array}{c} 
\bar{\myvv} \\ \myv 
\end{array}\bigg) 
= \bigg(\begin{array}{c} 
0 \\ 0
\end{array}\bigg),
\end{equation}
where $\bar{\myvv}$ and $\myv$ are quaternion valued fields.

With the stationary ansatz we have the eigenvalue equations
\begin{equation}\label{Dirac}
\bigg(
 \begin{array}{cc}
   0 & \slashed{D}^{\dag} \\
\slashed{D} & 0
\end{array} \bigg)
\bigg(\begin{array}{c} 
\bar{\myvv} \\ \myv 
\end{array}\bigg) 
= \en\bigg(\begin{array}{c} 
\bar{\myvv} \\ \myv 
\end{array}\bigg). 
\end{equation}
This equation is closely related to a Dirac equation  studied by Jackiw and Rebbi in  \cite{JacReb:75},
where they  investigated the zero-energy solutions. Bais
and Troost \cite{BaiTro:81} looked at the same Dirac equation, extending
the analysis to higher energies, and looking for bound states. 
 
The other second order equation which it is possible to arrange (\ref{Dirac})
into is 
\begin{equation}\label{vmaster}
\slashed{D}\slashed{D}^{\dag}\myv=\en^2 \myv.
\end{equation}
This equation is equivalent to \eqref{master}, provided $\omega \neq 0$. The map between solutions 
is 
\begin{equation}
\label{susytrafo}
 \slashed{D}\bar{\myvv}=\en \myv,\quad \text{and}  \quad  \slashed{D}^\dagger \myv=\en \bar{\myvv},
\end{equation}
 as seen in (\ref{Dirac}).  Explicitly, we can obtain a solution of \eqref{vmaster} from a solution of \eqref{master}
by  applying $\slashed{D}$ to each side of (\ref{master}) and using \eqref{susytrafo}:
\begin{equation} 
\slashed{D}\slashed{D}^{\dag}\slashed{D} \bar{\myvv}=\en^2
\slashed{D}\bar{\myvv} \Rightarrow \slashed{D}\slashed{D}^{\dag} \myv = \en^2  \myv,
\end{equation}
where we also divided by $\en$.  Conversely, we can also map a solution of \eqref{vmaster} into a solution of \eqref{master} by substituting 
for $\myv$ using  \eqref{susytrafo},  provided $\en\neq 0$.
 As we shall see later in this paper,  it is fruitful to investigate both (\ref{master}) and (\ref{vmaster}), using the
map \eqref{susytrafo} to relate the results.

Finally, we note that the transformation \eqref{susytrafo} also provides a convenient way of checking if a solution $\bar{\myvv}$ of \eqref{master} satisfies the background gauge condition.  According to \eqref{QLBEexpression},  the latter is the requirement that the real  part (in the quaternionic sense) of $ \slashed{D}\bar{\myvv}$  vanishes. In other words, to see if  the background gauge condition holds we simply check if the quaternion $\myv$ obtained from  $\bar{\myvv}$ according to \eqref{susytrafo} has a vanishing real part.

\section{The quaternionic wave equation in the background of the BPS monopole}

\subsection{Structure and symmetries of the wave equation}
We  begin our detailed study of the equation   (\ref{master}) in  the case where the background 
field  is the BPS  monopole \eqref{hedgehog}, with the profile functions given in \eqref{WH}. The Dirac operator introduced in \eqref{Diracops}  now takes the form
\begin{equation}
\label{susygen}
 \slashed{D}^\dagger_{BPS} = (e_i\partial_i) +\frac {(1-W)}{r}(\ev\times\x)\cdot \tv  - \frac {H}{r}\x\cdot\tv. 
\end{equation}
Using the equations (\ref{WHeqns}) satisfied by the profile functions  \eqref{WH} we find
\begin{equation}\label{Diphi}
 \begin{split}
D_i\phi&=\left(\partial_i+\frac{x_k}{r^2}(1-W)\ep_{aik}t_a\right)\left(\frac{H}{r^2}x_bt_b\right) \\
       &=\delta_{ib}\frac{H}{r^2}t_b+\frac{x_ix_b}{r}\left(\frac{H}{r^2}\right)'t_b
           +(1-W)\left(\frac{H}{r^2}\right)\frac{x_kx_b}{r^2}\ep_{aik}\ep_{abc}t_c\\
       &=\frac{x_ix_a}{r^3}H't_a - \frac{x_ix_a}{r^2}(1+W)\frac{H}{r^2}t_a+\frac{WH}{r^2}t_i,
\end{split}
\end{equation}
so that 
\begin{equation}\label{waveop}
\begin{split}
(\slashed{D}^{\dag}\slashed{D})_{BPS} &=-D_i^2-\phi^2+2D_i\phi e_i  \\
&=-\Delta- \frac{2(1-W)}{r^2}\Lv\cdot \tv - \frac{(1-W)^2}{r^2}\tv^2+\frac{(1-W)^2-H^2}{r^2}(\x \cdot \tv)^2 \\
&\qquad+\frac{2H'}{r}(\x \cdot \tv)(\x \cdot \ev) -\frac{2(1+W)H}{r^2}(\x \cdot \tv)(\x\cdot \ev)+\frac{2WH}{r^2}(\ev \cdot \tv), 
\end{split}
\end{equation}
where the orbital angular momentum operator $\Lv$  has components
\begin{equation}
\label{angmom}
L_i=-\epsilon_{ijk} x_j \partial_k,
\end{equation}
and  the Laplace operator  can be written as 
\begin{equation}
 \label{threelaplace}
\Delta =  \partial_1^2 + \partial_2^2 + \partial_3^2 = \frac 1 r \partial_r^2 r  + \frac {1}{r^2}
\mathbf{L}^2.
\end{equation}
We  also note that  the  differential operator  in \eqref{vmaster}  now takes the form
\begin{equation}
 \label{vwaveop}
\begin{split}
 (\slashed{D}\slashed{D}^{\dag})_{BPS}&=-D_i^2-\phi^2 \\
&=-\Delta- \frac{2(1-W)}{r^2}\Lv\cdot \tv - \frac{(1-W)^2}{r^2}\tv^2+\frac{(1-W)^2-H^2}{r^2}(\x\cdot \tv)^2. 
\end{split}
\end{equation}

We will use the quaternionic formulation for studying the operators in \eqref{waveop} and \eqref{vwaveop}.  This means that we will  let them  act on  functions 
\begin{equation}
\label{quatfct}
 q:\RR^3 \rightarrow \HH\otimes\mathfrak{su}_2, 
\end{equation}
as explained  after \eqref{Diracops}.
We begin with \eqref{masterr}, which arises directly from the linearisation discussed in the previous section,  and  look for eigenfunctions, i.e. solutions of 
\begin{equation}
 \label{masterr}
(\slashed{D}^{\dag}\slashed{D})_{BPS} \bar q = \omega^2 \bar q.
\end{equation}

The BPS monopole is spherically symmetric  in the sense that a spatial rotation can be compensated for by an iso-rotation. 
The operator generating the combined spatial and iso-rotations can be expressed in terms of the angular momentum operator $\mathbf{L}$ \eqref{angmom}, the spin operator $\sv =  \frac{1}{2}\ev $ (whose components  act on the 
quaternion part of   \eqref{quatfct} via commutator), and the  isospin operator $\tv$ (whose components $t_i=-\frac{i}{2}\tau_a$    act in the adjoint representation on the  $\mathfrak{su}_2 $ part of $q$). 
It is easy to check  that, with our conventions,  the components of the generalised angular
momentum operator 
\begin{equation}\label{JeqLst} 
\mathbf{J} = \Lv + \sv +\tv
\end{equation}
 satisfies $[J_i,J_j]=\epsilon_{ijk}J_k$ and  commutes with $\slashed{D}_{BPS}$ and $ \slashed{D}^{\dag}_{BPS}$  defined in \eqref{susygen}. As a result we are able to organise eigenfunctions of the operator $(\slashed{D}^{\dag}\slashed{D})_{BPS}$ in terms of multiplets of the generalised angular momentum operator $\mathbf{J}$.
Doing this in practice is the subject of the next section.

\subsection{Partial wave analysis}
\label{partialsect}

In order to split the set of eigenfunctions in \eqref{masterr} into  irreducible representations (multiplets) of the generalised angular momentum operator $\mathbf{J}$ we apply basic results from the representation theory of  $SU(2)$. Denoting irreducible representations of $SU(2)$ by their spin $j\in \frac  1 2 \NN$  we recall the basic tensor product decomposition rule
\begin{equation}
 \label{oldhat}
j_1\otimes j_2 =\bigoplus_{n=|j_1-j_2|}^{j_1+j_2} n.
\end{equation}
From the point of view of representation theory, we may think of quaternions as a direct sum of a spin $0$ and a spin $1$ representation of $\sv$:
\[
\HH_1\simeq 0\oplus 1.
\]
The quaternionic function \eqref{quatfct} can be viewed as a tensor product of $(i)$ a scalar function on $\RR^3$, $(ii)$ a quaternion and,  $(iii)$ a spin $1$ representation of the isospin operator $\tv$. The three terms in   $\mathbf{J}$  act on each of these separately according to  $\mathbf{J} =\Lv\otimes 1\otimes 1 +
1\otimes\sv\otimes 1 + 1\otimes 1 \otimes\tv$. If we split the space of scalar functions on $\RR^3$ further into a tensor product of functions of the radial coordinate $r$ and the space $L^2(S^2)$ of  square-integrable functions  on the two-sphere,  then $\mathbf{J}$
acts trivially on the radial functions, and  the decomposition of $L^2(S^2)$ into irreducible representations of the orbital angular momentum operator $\Lv$ is the usual decomposition of functions on the two-sphere  into spherical harmonics:
\[
 L^2(S^2) =\bigoplus_{l=0}^\infty l.
\]
Thus, applying the rule \eqref{oldhat} to the tensor product of quaternions and $\mathfrak{su}_2$ we observe
\[
 (0\oplus 1)\otimes 1 = 1\oplus 0 \oplus 1 \oplus 2.
\]
Tensoring further with irreducible representations  of the orbital angular momentum operator  $\Lv$ we deduce that for $l\geq 2$, 
 \begin{align}\label{genirreps1}
l \otimes  (0\oplus 1)\otimes 1 &= l\otimes (0\oplus 1  \oplus 1 \oplus 2)\nonumber \\
&=l\oplus(l-1)\oplus l \oplus (l+1 ) \oplus(l-1)\oplus l \oplus (l+1 )\nonumber \\
&\quad  \oplus (l-2) \oplus (l-1) \oplus l \oplus (l+1) \oplus (l+2).
\end{align}
For $ l=1$ we have
 \begin{align}\label{genirreps2}
1 \otimes  (0\oplus 1)\otimes 1 &= 1\otimes (0\oplus 1  \oplus 1 \oplus 2)\nonumber \\
&=1\oplus 0 \oplus 1 \oplus 2 \oplus 0 \oplus 1 \oplus  2  \oplus 1  \oplus 2  \oplus 3, 
\end{align}
while for $l=0$ we have 
 \begin{align}\label{genirreps3}
0 \otimes  (0\oplus 1)\otimes 1 &=0\oplus 1  \oplus 1 \oplus 2.
\end{align}

We can use these equations to count the number of angular momentum representations 
that can occur for a given value of the total angular momentum $j$. To do this, we fix a value of $j$ and count, with multiplicity, the values of $l$ which occur on the right hand side of equations \eqref{genirreps1} - \eqref{genirreps3}.  For $j\geq 2$ we find that $l=j$ contributes four times, $l=j-1$ and $l=j+1$ three times each, and $l=j-2$ and $l=j+2$ once each, giving a total of 12 modes. For $j=1$, the value  $l=j+2=3$ does not contribute, and $l=j=2=-1$ is impossible, so only 10 modes occur. Finally, for $j=0$, the possibility $l=0$ occurs once, the possibility $l=1$ twice and $l=2$ once, giving a total of four modes. For a given value of $j$,  each of the modes has the usual, additional  degeneracy  of $2j+1$.  However, because of overall invariance of the situation under generalised rotations, these $2j+1$ states  are physically equivalent (and  obey the same differential equation).

\subsection{The $j=0$ sector}

The four modes with  $j=0$  can be constructed very simply by combining $l=0,1,2$ functions with the quaternionic and isospin degrees of freedom to obtain overall scalars under the action of $\mathbf{J}$. Since all states must have isospin 1, 
the $l=0$ (constant) function can only be combined  with $\ev \cdot \tv$ to obtain an overall scalar. 
 Using the cartesian coordinates of the unit vector $\x$ on the spatial two-sphere as the three   $l=1$ states, we can obtain one overall scalar from the  scalar ($s=0$) field as $ \x\cdot \tv$. Another overall scalar can be constructed from the vector $(s=1)$ field as $\x\cdot(\ev \times \tv) $. Finally, the five independent functions spanning the $l=2$  multiplet are the components of the tensor $
\hat x_i\hat x_j  - \frac 1 3\delta_{ij}$. We can combine these with the isovector and the spin $1$ part of the quaternion as
$(\x\cdot \tv)(\x \cdot \ev)-\frac{1}{3}\ev\cdot \tv$ to obtain an overall scalar.  We thus have four basis states of the  $j=0$ sector, 
and can use any linear combination (with coefficients being functions of $r$)   to study the $j=0$ sector of  \eqref{masterr}. It turns out that, 
for our purposes, the following basis states are most convenient:
\begin{subequations}\label{zeromodes}
\begin{align}
v_1&=(\x\cd \tv)(\x \cd \ev)-\ev\cd \tv\\
v_2&=(\x\cd \tv)(\x\cd \ev)\\
v_3&=\x\cd (\ev \times \tv)\\
v_4&=\x\cd \tv.
\end{align}
\end{subequations}
We note that $v_3$ and $v_4$
are perturbations of the hedgehog fields as defined in
(\ref{hedgehog}) and are the assumed shape of the perturbations used
in \cite{ForVol:03}. 

It is worth interpreting our chosen basis physically before proceeding with the mathematical analysis. The generator $\x\cd \tv$  in the isospin Lie algebra is in the direction of the asymptotic Higgs field and thus the  generator of the unbroken $U(1)$ subgroup of the original isospin symmetry. Even though the YMH model studied here is not a realistic physical model, we adopt a terminology where  this $U(1)$ is interpreted as the gauge group of electromagnetism. Then we note that $v_2$ is the unbroken part of the $\mathfrak{su}_2$ gauge field and can therefore be thought of as the photon field. The other spin $1$ states $v_1$ and $v_3$ are eigenstates of $(\x\cd\tv)^2$ with eigenvalue $-1$ and therefore have electric charge $\pm 1$. Neither of them is an eigenstate of $\x\cd\tv$ so both should be thought of as linear combinations of excitations of  the charged $W$-bosons  in this model. Finally, the spin zero state $v_4$ is proportional to the asymptotic value of  Higgs field and describes a 
massless and uncharged excitations of the Higgs field.  We still need to ascertain which linear combination of these states satisfies the background gauge condition \eqref{backgd_gauge}. We will do this below by applying the operator $\slashed{D}$, as outlined at the end of Sect.~\ref{BPSTimeDependent}.

Our next task is the application  of the operator $(\slashed{D}^{\dag}\slashed{D})_{BPS}$
 (\ref{waveop})  to a linear combination
\begin{equation}
\bar{\myvv}=\sum_{n=1}^4 \basis_n(r)v_n,
\end{equation}
of the basis functions of the $j=0$ sector.  
In order to organise the calculation we note that $(\slashed{D}^{\dag}\slashed{D})_{BPS}$
is a linear combination 
of  the operators $\Delta$, $\Lv\cdot \tv$, 
$\tv^2$, $(\x\cdot \tv)^2$, $ (\x \cdot \ev)(\x\cdot \tv)$ and $\ev \cdot \tv$, 
multiplied by simple functions of $W$ and $H$.  We list, in 
 Table \ref{tableops}, 
the action of the relevant  operators on the  modes $v_1,\ldots,v_4$  (\ref{zeromodes}). 

\begin{table}\label{tableops}
\caption{The action of the operators in (\ref{waveop}) on each of the zero
  angular momentum modes (\ref{zeromodes}).}
\centering
\begin{tabular}{|c|c|c|c|c|c|c|}
 \hline &$v_1$&$v_2$&$v_3$&$v_4$ \\ \hline
$\Lv^2$&$-2(v_1+2v_2)$&$-2(v_1+2v_2)$&$-2v_3$&$-2v_4$\\
$\Lv \cdot \tv$&$v_1+2v_2$&$v_1+2v_2$&$v_3$&$2v_4$\\
$\tv^2$&$-2v_1$&$-2v_2$&$-2v_3$&$-2v_4$\\
$(\x \cdot \tv)^2$&$-v_1$&$0$&$-v_3$&$0$\\
$(\x \cdot \ev)(\x \cdot \tv)$&$v_1$&$0$&$v_3$&$0$\\
$\ev \cdot \tv$&$v_1-2v_2$&$-v_1$&$v_3-2v_4$&$-v_3$\\
 \hline 
\end{tabular}
\end{table}

 Using   (\ref{tableops}),   we find that  (\ref{masterr}) becomes
\begin{equation}\label{explicitmaster}
\begin{split}
&\Big( - r(r\basis_1)''v_1 -r(r\basis_2)''v_2 - r(r\basis_3)''v_3  -r(r\basis_4)''v_4 \Big) 
+ \Big(2\basis_1v_1+2\basis_2v_1  +4\basis_1v_2\\
&\quad +4\basis_2v_2 + 2\basis_3v_3   +
2\basis_4v_4 \Big) 
+\Big(-2(1-W)\basis_1v_1 -2(1-W)\basis_2v_1 \\ 
&\quad-4(1-W)\basis_1v_2-4(1-W)\basis_2v_2
  - 2(1-W)\basis_3v_3 - 4(1-W)\basis_4v_4 \Big) \\
&\quad +\Big(2(1-W)^2\basis_1v_1 +2(1-W)^2\basis_2v_2 + 2(1-W)^2\basis_3v_3  + 2(1-W)^2\basis_4v_4\Big) \\
&\quad  +\Big( (-(1-W)^2+H^2)\basis_1v_1  
+ (-(1-W)^2+H^2)\basis_3v_3 \Big) \\
&\quad+\Big( 2(rH'-(1+W)H)\basis_1v_1 + 2(rH'-(1+W)H)\basis_3v_3 \Big)  
 + \Big(2WH\basis_1v_1\\
&\quad-2WH\basis_2v_1 -4WH\basis_1v_2+2WH\basis_3v_3 - 2WH\basis_4v_3 -4WH\basis_3v_4 \Big) \\
&=r^2\omega^2(\basis_1v_1+\basis_2v_2+\basis_3v_3+\basis_4v_4). 
\end{split}
\end{equation}
Each group of terms which is bracketed in (\ref{explicitmaster}) comes from one part of (\ref{masterr}).
We compare coefficients of $v_1, v_2,
v_3$ and $v_4$ in (\ref{explicitmaster}) to obtain equations for the radial functions
$\basis_n(r)$.
 We find that $\basis_1(r)$ and $\basis_2(r)$ are coupled  and that 
 $\basis_3(r)$ and $\basis_4(r)$ are coupled. As noted previously,
 $\basis_3(r)$ and $\basis_4(r)$ are related to the functions $w(r)$ and $h(r)$ in
 \cite{ForVol:03}, as they are also perturbations of the basic hedgehog fields
 (\ref{WH}). The coupling is not quite
 symmetric  but this can be remedied by 
defining 
\begin{equation}
w=r\basis_3, \quad h=\frac{r\basis_4}{\sqrt{2}}.
\end{equation}
A similar redefinition 
\begin{equation}
 v=r\basis_1,\quad 
\alpha=\frac{r\basis_2}{\sqrt{2}},
\end{equation}
for the other coupled system aids comparison with the previous work, and
again results in the coupling being symmetric. 

With  these  abbreviations, we conclude that the insertion of 
\begin{equation}
\label{masterexpansion}
 \bar{\myvv}=\frac  1 r (  v  v_1 + \sqrt{2} \alpha   v_2 + w  v_3 + \sqrt{2} h  v_4),
\end{equation}
into \eqref{masterr} gives a system of second order differential equations which decouples into two systems:
 \begin{subequations}
\label{bkchannels}
 \begin{align}
 \left( -\frac{d^2}{dr^2}+\frac{3W^2+H^2-1}{r^2}  \right) v -\frac{2\sqrt{2}W(H-1)} {r^2} \alpha &= \en^2 v , \label{bkv}\\
 \left( -\frac{d^2} {dr^2}+\frac{2W^2+2} {r^2}  \right) \alpha - \frac{ 2\sqrt{2}W(H-1) } {r^2}  v &=\en^2 \alpha,     \label{bka}
 \end{align}
 \end{subequations}
and
\begin{subequations}\label{fvchannels}
 \begin{align}  
\left(-\frac{d^2}{dr^2}+\frac{3W^2+H^2-1}{r^2} \right) w-\frac{2\sqrt{2}WH}{r^2} h &=\en^2w , \label{fvw}\\
\left(-\frac{d^2}{dr^2}+\frac{2W^2}{r^2} \right) h-\frac{2\sqrt{2}WH}{r^2} w &=\en^2 h. \label{fvh}
\end{align}  
\end{subequations}
As noted,  $w(r)$ and $h(r)$ are perturbations of the hedgehog
fields $W(r)$ and $H(r)$, and thus we could have obtained (\ref{fvchannels})
by linearising (\ref{WHeqns}). This is what is done in  \cite{ForVol:03}. In the following sections we will investigate bound and scattering states in the  systems 
\eqref{bkchannels} and \eqref{fvchannels}. Recalling the interpretation  of the basis \eqref{zeromodes} we note that the  system \eqref{bkchannels}  describes a photon mode interacting with a  $W$-boson mode, and that  \eqref{fvchannels} describes 
a   massless Higgs perturbation interacting with a $W$-boson mode. Correspondingly, we will often refer to the former as the photon system and the latter as the Higgs system.

So far, we have focussed  on the 
 equation \eqref{master}  since it is directly related to the linearised  YMH equations. The 
 alternative but equivalent equation \eqref{vmaster}, however, also   plays an important role in our analysis. 
 As explained at  the end of Sect.~\ref{BPSTimeDependent},  the  background gauge condition \ref{backgd_gauge}  is conveniently implemented in this formulation. It also turns out that the equation\eqref{vmaster} is also sometimes easier to analyse than \eqref{master}.
 
Recalling the definition \eqref{vwaveop}, we thus consider the eigenvalue problem
\begin{equation}
 \label{vmasterr}
(\slashed{D}\slashed{D}^{\dag})_{BPS} \myv=\en^2 \myv.
\end{equation}
Inserting, in analogy to \eqref{masterexpansion}, the expression
\begin{equation}
\label{vmasterexpansion}
 \myv=\frac  1 r (  \chi  v_1 + \sqrt{2} \psi   v_2 + \xi  v_3 + \sqrt{2} \zeta v_4),
\end{equation}
for four radial functions $\xi,\eta, \chi,\psi$ 
we obtain  again two systems of second order differential equations:
\begin{subequations}
\label{xizetachannels}
\begin{align}
\left(-\frac{d^2}{dr^2}+\frac{W^2+H^2+1}{r^2}\right)\xi &= \omega^2\xi,
\label{xi} \\
\left(-\frac{d^2}{dr^2}+\frac{2W^2}{r^2}\right)\zeta&=\omega^2\zeta ,\label{zeta}
\end{align}
\end{subequations}
and
\begin{subequations}
\label{psichichannels}
\begin{align}
\left(-\frac{d^2}{dr^2}+\frac{W^2+H^2+1}{r^2}\right)\chi +\frac{2\sqrt{2}W}{r^2}\psi&=
 \omega^2\chi \label{chi},\\
\left(-\frac{d^2}{dr^2}+\frac{2W^2+2}{r^2}\right)\psi+\frac{2\sqrt{2}W}{r^2}\chi&=
\omega^2\psi. \label{psi}
\end{align}
\end{subequations}
According to \eqref{susytrafo}, we can relate a  solution \eqref{masterexpansion} of \eqref{masterr}  to  a solution \eqref{vmasterexpansion} of \eqref{vmasterr}  via the Dirac operator $\slashed{D}^\dagger_{BPS}$ in  \eqref{susygen}
according to
\begin{equation}
\label{susymap}
 \omega \bar{ \myvv} = \slashed{D}^\dagger _{BPS}  \myv.
\end{equation}
We are now in a position to apply  the background gauge condition \eqref{backgd_gauge} to solutions of \eqref{vmaster}. As discussed  at the end of Sect.~\ref{BPSTimeDependent},  we can do this by ensuring that the  quaternion $\myv$ has no real quaternionic part. Of all the basis elements \eqref{zeromodes}, only $v_4$ is real in the quaternionic sense. Hence solutions of the form \eqref{vmasterexpansion} correspond, via \eqref{susymap}, to solutions of the linearised YMH equations if the coefficient function $\zeta$ in \eqref{vmasterexpansion} is identically zero.  Note that, according to \eqref{xizetachannels}, this can be imposed consistently since $\zeta$ satisfies a homogeneous linear equation and does not couple to any other mode.

Working out the relation \eqref{susymap} explicitly is a little tedious, but can be done by careful application of  some of the results in  \eqref{tableops} as well as the application of  the operator $(e_i\partial_i)$ to $v_1,\ldots,v_4$.    
It can be seen without too much effort that the transformation \eqref{susymap}  permutes the basis elements $v_1, \ldots,v_4$.
In particular,  the coefficient functions $\xi,\zeta$  of   $v_3,v_4$ are mapped to  the coefficient functions $ v,\alpha $ of  $v_1,v_2$.
Equally,   the coefficient functions $\chi,\psi$ of $v_1,v_2$ are mapped to  the coefficient functions  $w,h$ of $v_3,v_4$.  Since the background gauge condition does not  restrict the functions $\chi$ and $\psi$, we deduce that any solution of \eqref{fvchannels}  satisfies the background gauge condition.

However, in order to understand the implication of the background gauge for \eqref{bkchannels}, we need to know the relation between $\xi,\zeta$   and  $ v,\alpha $. 
To compute the effect of applying \eqref{susygen} to the basis \eqref{zeromodes} we note that, in terms of the usual spherical  coordinates  $(r,\theta,\varphi)$ for $\RR^3$, 
\[
\x = \begin{pmatrix} \sin\theta \cos \varphi \\ \sin\theta \sin\varphi \\ \cos \theta\end{pmatrix},\qquad
\hat{\mathbf{\theta}} = \partial_{\theta} \x =  \begin{pmatrix} \cos \theta \cos \varphi \\ \cos\theta \sin\varphi \\ -\sin \theta\end{pmatrix},\qquad
\hat{\mathbf{ \varphi}} =  \partial_{\varphi}\x  =\begin{pmatrix} - \sin\varphi \\ \phantom{-} \cos \varphi \\ 0 \end{pmatrix},\qquad
\]
and 
\[
\x\times \mathbf{L}=\hat{\mathbf{\theta}} \partial_{\theta} + \frac{\hat {\mathbf{ \varphi}} }{\sin\theta}\partial_\varphi.
\]
Therefore
\[
 e_i\partial_i  = \x\cd\ev\,\partial_r + \frac{1}{r}\hat{ \mathbf{\theta} } \cd\ev\,\partial_{\theta} +
\frac{1}{r\sin\theta}\hat{\mathbf{\varphi}}\cd \ev \,\partial_\varphi = \x\cd\ev\,\partial_r +\frac{1}{r}\ev\cd \x\times\mathbf{L}.
\]
Then we compute, for example, 
\[
 (e_i\partial_i) \, v_4 =  -\frac 1 r  v_1,   \quad  (e_i\partial_i)\, v_3=\frac 1 r (2v_2-v_1).
\]
Evaluating the other terms  \eqref{susymap} we arrive at 
 \begin{subequations}
 \label{trans}
\begin{align}  
 \omega\alpha&=\frac{d\zeta}{dr} -\frac{\zeta}{r}+ \frac{ \sqrt{2} W} {r} \,\xi, \label{tra}\\
 \omega v&=-\frac{d\xi }{dr}  -  \frac{H}{r}\,\xi - \frac{\sqrt{2} W}{r} \zeta. \label{trv}
\end{align}
 \end{subequations} 
Remarkably, the transformation \eqref{trans} relates the coupled system \eqref{bkchannels} to the decoupled system \eqref{xizetachannels}. Thus, the most efficient way of studying the  the coupled system \eqref{bkchannels} is to look at \eqref{xizetachannels} instead, and then apply \eqref{trans}. In addition, if one is only interested in bosonic modes satisfying the background gauge condition (as we are in this paper), one should set $\zeta$ to zero.

\section{Spectral properties of the  $j=0$ sector}
\label{specsec}
\subsection{General considerations}
\label{qualispec}
Before we turn to the numerical investigation of bound states and scattering in the $j=0$ sector, we note some general features of  the systems
 \eqref{bkchannels} and \eqref{fvchannels} on the one hand and \eqref{xizetachannels} and \eqref{psichichannels} on the other.  For this purpose we note 
the asymptotic behaviour of the coefficient functions  $W$ and $H$ \eqref{WH}:
\begin{align}
\label{zeroasy}
W(r) \approx1 - \frac{r^2}{6}, \quad 
H(r) \approx -\frac{r^2}{3} \quad \text{for small} \, r,
\end{align}
and 
\label{WHasympt}
 \begin{align}
W(r) \approx \frac{r}{2}e^{-r} ,\quad  H(r) \approx  1-r + O(e^{-r})  \quad \text{for large} \, r.
\end{align}

It follows that the systems  \eqref{bkchannels}  and  \eqref{fvchannels}   both decouple at large values of $r$. We can think of them as two channels which 
are coupled in the region of the background monopole but whose coupling falls off exponentially fast as we go away from the core of the monopole. 
It is instructive to set the coupling terms to zero and consider the resulting single channels. The equations \eqref{bkv} for $v$ and \eqref{fvw} for $w$ 
are the same after decoupling, and take the form
\begin{equation}\label{bkapprox}
\left(-\frac{d^2}{dr^2}+\frac{3W^2+H^2-1}{r^2}\right)v=\omega^2v.
\end{equation} 
The potential  appearing in this equation has the asymptotic form
\begin{equation}
\label{potasy}
 \frac{3W^2+H^2-1}{r^2} \approx 1-\frac{2}{r} + O(e^{-r}) \quad \text{as} \; r \rightarrow \infty,
\end{equation}
and, in particular, tends to the positive constant $1$  for large $r$.  Thus, 
thinking of \eqref{bkapprox} as the spatial part of a radial wave equation,  we  see that   its wave solutions correspond to massive particles of mass $1$, which is in agreement with our interpretation of $v$ and $w$ as  excitations of the $W$-bosons in YMH theory.  By contrast, the equation for  the function $\alpha$ 
after decoupling contains the potential 
\begin{equation}
\label{alphapot}
 \frac{2W^2 + 2} {r^2} \approx \frac{2}{r^2} + O(e^{-r}) \quad \text{for large} \, r,
\end{equation}
while the equation for  the function $h$ 
after decoupling contains the potential 
\begin{equation}
\label{hpot}
  \frac{2W^2} {r^2} \approx  O(e^{-r}) \quad \text{for large} \, r.
\end{equation}
Therefore, the corresponding radial waves are massless excitations, in agreement with their interpretation as, respectively, photon and Higgs excitations. 

\begin{figure}[htp]
\centering
\includegraphics[width=8truecm]{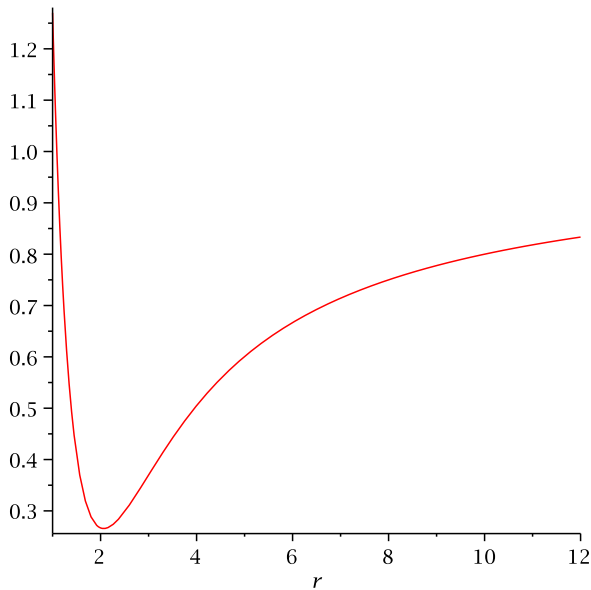}
\caption{The potential $\frac{3W^2+H^2-1}{r^2}$  plotted against $r$.}
\label{fig:pot}
\end{figure}

The potential $(3W^2+H^2-1)/r^2$ is plotted in Fig.~\ref{fig:pot}. The plot and the  appearance of the attractive  Coulomb tail in the asymptotic form \eqref{potasy}  suggest that the Sturm-Liouville problem \eqref{bkapprox} should have  an infinity of bound states for $\omega^2<1$ accumulating at
$\omega^2=1$, with a continuous spectrum for $\omega^2>1$.

The coupling of massless channels to a massive channel with infinitely many bound states constitutes the generic  situation for the occurrence of Feshbach resonances, or quasi-normal modes. We give a brief summary and  references  in Appendix~\ref{QNMTheory}. On the grounds of the general theory, we might expect both \eqref{bkchannels} and \eqref{fvchannels} to exhibit  resonance scattering 
for  $0<\omega^2<1$.  However, we  also note an important difference between the systems \eqref{bkchannels} and \eqref{fvchannels}. It follows from \eqref{zeroasy} that the coupling terms  in \eqref{bkchannels} are singular at $r=0$, whereas they are smooth for \eqref{fvchannels}.
 
Turning now to the systems   \eqref{xizetachannels} and \eqref{psichichannels} we observe they, too, consist of a  massive and a massless channel each. The potential in the massive channels is 
\begin{equation}
\label{singchanasy}
 \frac{W^2+H^2+1}{r^2} \approx
1-\frac{2}{r} + \frac{2}{r^2} + O(e^{-r})\quad \text{as} \;  r \to \infty, 
\end{equation}
and, because of the attractive Coulomb term,  should again support infinitely many bound states.
However, while the system  \eqref{psichichannels}  has the same generic form as  the photon  and Higgs systems \eqref{bkchannels} and \eqref{fvchannels}
discussed above, the equations in \eqref{xizetachannels} are decoupled. As a result, we expect this system to have true bound states for the excitation denoted $\xi$
 in the range $0<\omega^2<1$. There are scattering states in the excitation $\zeta$, but these do not satisfy the background gauge condition, as already discussed after \eqref{susymap}.   Hence,  on the basis of the   decoupled form \eqref{xizetachannels},  we expect  the system \eqref{bkchannels} to have  infinitely many bound states and no (physical) scattering states in the range $0<\omega^2<1$. This is in marked contrast with the Feshbach resonances of the Higgs system \eqref{fvchannels}.

Before we present our numerical results we briefly comment 
on or numerical methods. 
All of our scattering calculations  require  a shooting-to-a-fitting-point method for finding
solutions which decay at large enough values in  one of the channels. Numerical instability would
always create difficulties where the exponentially increasing term creeps into
the solution. Thus, we  choose
an appropriately large value of $r$ to integrate to and  use a  shooting method to find the decaying solution.

\subsection{Bound states}

It is well known that the zero-modes of the 't Hooft-Polyakov monopole give rise to zero-energy bound states of the linearised YMH equations. 
The zero-modes of the 't Hooft-Polyakov monopole  are obtained from infinitesimal translations in $\RR^3$ and a special `large' gauge transformation generated by the Higgs field itself. This transformation does not vanish at infinity and is therefore considered as a physical symmetry transformation rather than a gauge transformation. 
The  infinitesimal form of the `large gauge
  transformation' generated by the Higgs field is 
\begin{equation}\label{magneticfield}
\delta\phi = 0 , \quad 
  \delta A_i =  D_i\phi=  -B_i.
\end{equation}
In our quaternionic formulation the corresponding zero-mode is simply 
 \begin{equation}
\label{zerom}
\myvv_\phi = e_iB_i,
\end{equation}
while the other three zero-modes related to translations are 
\begin{equation}
 \label{zeromk}
 \myvv_k = e_k(e_iB_i).
\end{equation}
It is easy to check that  \eqref{zerom} (and hence \eqref{zeromk}) satisfy the linearised Bogomol'nyi equation \eqref{QLBE}, and hence \eqref{masterr} for $\omega=0$. 
Of the four zero-modes just found, only 
 $\myvv_\phi$ has $j=0$. It has the explicit form
\begin{equation}
B_i=-D_i\phi
=-\frac{x_ix_a}{r^3}H't_a + \frac{x_ix_a}{r^2}(1+W)\frac{H}{r^2}t_a-\frac{WH}{r^2}t_i,
\end{equation}
and is thus a linear combination of the basis states $v_1$ and $v_2$ as defined in
(\ref{zeromodes}). In particular,  it is   therefore a solution of the photon system  \eqref{fvchannels}.

Bais and Troost first investigated  bound states in single channels arising  in the linearised BPS system in  \cite{BaiTro:81}. In particular,
 they  computed bound state energies in the system \eqref{xi}. For completeness of our discussion  we 
have repeated the numerical analysis here.  
Using the NAG shooting method \nag{D02KEF} for Sturm-Liouville type problems we find the
eigenvalues  $\omega^2_n$ and  hence  $\omega_n>0$ of the first few bound states of (\ref{xi}), to 3 significant figures. As explained in our 
qualitative discussion in the previous section, we expect there to be infinitely many Coulomb bound states
in this channel.  We can estimate their energies by neglecting the exponentially small terms in  the potential \eqref{singchanasy}
and using the standard formula for Coulomb bound state energies. In Table~\ref{table:xievalues}   we list both the 
numerically computed bound state energies of (\ref{xi}) and the Coulomb approximations 
\begin{equation}
\label{Omdef}
 \Omega_n=\sqrt{1 - \frac{1}{(n+1)^2}},
\end{equation}
where we used the standard expression for Coulomb bound state energies, recalling that $l=1$  in this case by virtue of the $1/r^2$-term in \eqref{singchanasy}.
Even for very small $n$, the Coulomb approximation is   surprisingly good.

\begin{table}[htp]
\centering
\[
\begin{array}{ccccccc}
 \hline \hline n&1&2&3&4&11  \\ \hline
\omega_n&0.877&0.946&0.970&0.980&0.997\\
\Omega_n&0.866&0.943&0.968&0.980&0.996\\
 \hline \hline
\end{array}
\]
\label{table:xievalues}
\caption{Values for $\omega_n$  for the eigenvalue problem (\ref{xi}), and $\Omega_n$  given in \eqref{Omdef}}
\end{table}

We have also computed the wavefunctions $\xi_n$, $n=1,2,\ldots $  for  each of the bound states. They have the standard form of Coulomb  bound states,   but it 
 is interesting to note that  they correspond, via  (\ref{trans}),  to  bound states in the coupled system \eqref{bkchannels}.  The functions $v$ and $\alpha$  for each of the bound states describe, via \eqref{masterexpansion}, the profile  of the  excited monopole. Since $h$ and $w$ both vanish for this excitation the bound state only involves the photon excitation $\alpha$ and  the  component $v$  of the field describing the  W-boson.

From the point of view of  the system \eqref{bkchannels}, it is surprising that there are true bound states. 
As explained in Sect.~\ref{qualispec}, one would generically
 expect two-channel problems like  \eqref{bkchannels} to have Feshbach resonances, but no bound states.

\subsection{Scattering}
\label{scatsec}

In Appendix \ref{QNMTheory} we discuss what kind of coupled systems are likely
to possess Feshbach resonances. Both (\ref{bkchannels}) and (\ref{fvchannels}) have the required features in the  parameter range  $0<\omega^2<1$:
 they consist of  two channels which  are weakly coupled at large distances, and are such that, after removing the coupling term, there are bound states
in one channel and  only scattering states  in the other. In fact,  the system
(\ref{fvchannels}) shows exactly the  expected Feshbach resonance behaviour.  This was first noticed by   Forg\'acs and Volkov in \cite{ForVol:03}, and we will revisit it  below. 
However, we will also see that the  system \eqref{bkchannels}, whose bound states we analysed in the previous section,  has no scattering states which satisfy the background gauge condition.

We begin with a brief review of the results of \cite{ForVol:03}. This will establish our conventions and terminology and serve as a preparation of our generalisation in Sect.~\ref{nonBPSsect}.
The system \eqref{fvchannels}  has a regular singular point at the origin, which means that the
numerical integration must start a little away from $r=0$. We make a series expansion near $r=0$ and  find the following leading terms 
\begin{equation}
  w \approx A r^2, \qquad h \approx B r^2 \quad  \text{for small} \;  r,
\end{equation}
 with real constants $A,B$. Because of the linearity of the problem we can scale the solution to fix one of those constants.
The remaining free constant plays the role of the shooting parameter. 

Once the  correct initial conditions are  determined to ensure a decaying solution $w$ (for $\omega^2 < 1$), 
we can consider the scattering problems in the massless channel  (\ref{fvh}). It has the general  form  
\begin{equation}\label{general}
\left(-\frac{d^2}{dr^2} + \frac{l(l+1)}{r^2} + V(r) - \omega^2\right) u (r) = 0,
\end{equation}
where $V(r)\to 0$ as $r \to \infty$ exponentially fast.  In the  equation for $h$, the total potential is \eqref{hpot}  so $l=0$. 

For large values of $r$, where $V$ can be neglected, the solution must be  a combination of  spherical  Bessel  functions
\begin{equation}\label{mpbesselcomparison}
u(r)=r\left(A_lj_l(\omega r) + B_ly_l(\omega r)\right), \quad  \text{for} \quad r \; \text{large}.
\end{equation}
Since,  asymptotically, $j_l(\omega r)\approx \sin(\omega r -\frac l 2 \pi)/(\omega r)$ and 
$y_l(\omega r)\approx -\cos(\omega r -\frac l 2 \pi)/(\omega r)$
we have the  asymptotic form
\[
u(r)\approx \frac{ \sqrt{A_l^2 +B_l^2} } {\omega^2} \sin(\omega r + \delta_l -\frac l 2 \pi) 
\]
of the radial wavefunction, with the phase shift  defined via
\[
\tan({\delta_l}) = - \frac {B_l}{A_l}.
\]
By evaluating both sides  of \eqref{mpbesselcomparison} for two large values of $r$ (or  by evaluating  both sides and their derivatives for large $r$) we  extract the coefficients  $A_l$ and $B_l$,  for each value of $\omega$ , and hence the phase shift $\delta_l$ at $\omega$. The latter  determines the partial  scattering cross section  according to the standard expression
\[
 \sigma_l(\omega) = \frac{4\pi(2l+1)} {\omega^2} \sin^2\delta_l(\omega),
\]
and the total cross section according to
\[
\sigma(\omega)=\frac{4\pi}{\omega^2}\sum\limits_{l=0}^{\infty}(2l+1)\sin^2\delta_{l}(\omega).
\]
  Near a  resonance, $\delta_l$  increases  rapidly by $\pi$. The function  $\sin^2\delta_l(\omega)$  takes values between $0$ and $1$, is maximal at the resonance and is thus an expedient quantity to plot when looking for resonances.

\begin{figure}[htp]
\centering
\includegraphics*[width=8truecm]{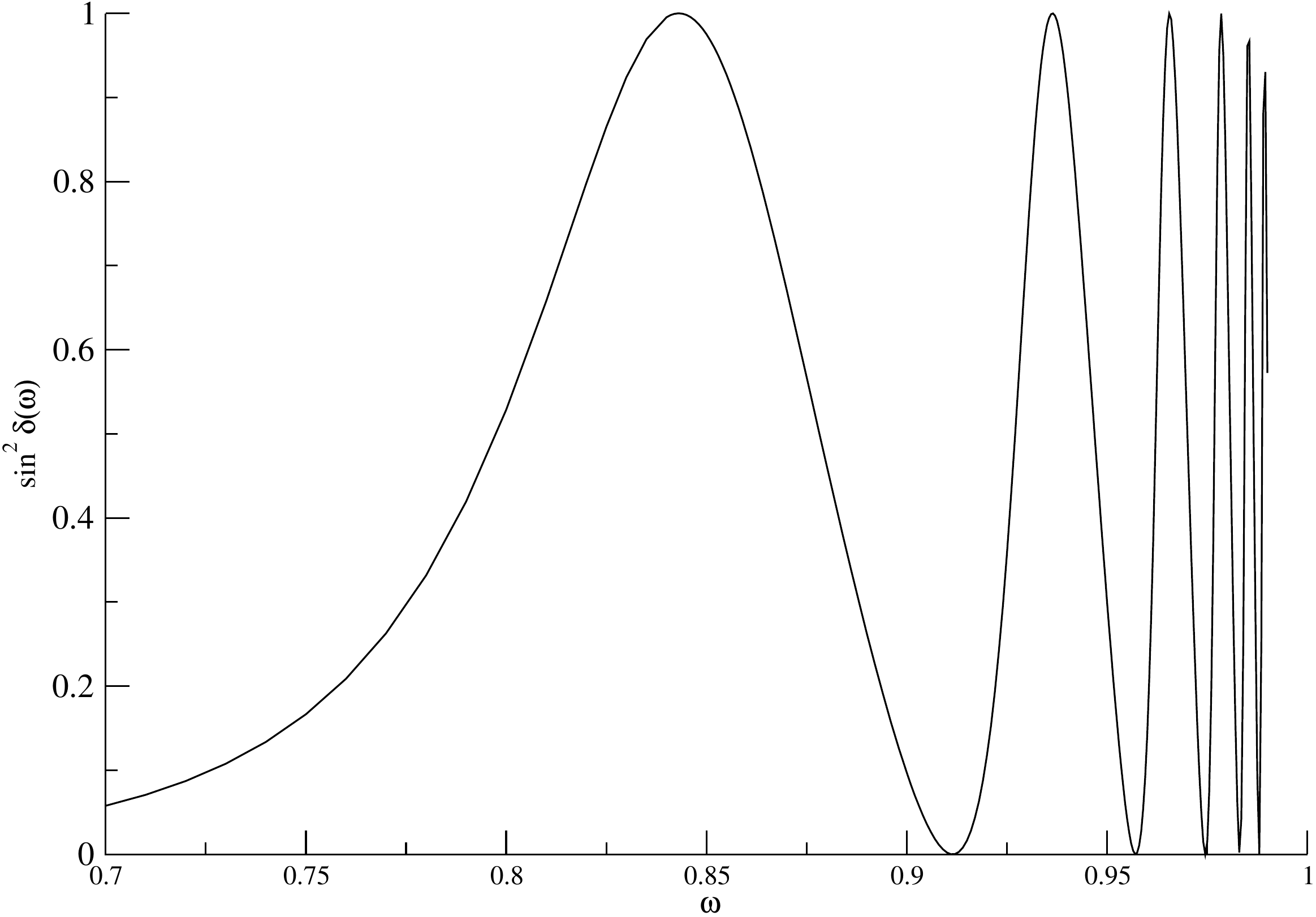}
\caption{For the system (\ref{fvchannels}), the partial scattering cross section
  $\sin^2\delta_0(\omega)$ is plotted against $\omega$.}
\label{fig:hscatxsec}
\end{figure}

Applying this procedure to   the $l=0$ contribution to the scattering cross section for the
massless channel $h$ from the Higgs  system (\ref{fvchannels})  we 
confirm the result shown in Fig.1 of \cite{ForVol:03},  suggesting infinitely many 
resonances as the energy approaches the critical value $\omega^2=1$. A graph
of $\sin^2\delta(\omega)$  against $\omega$ for the massless channel $h$ is
shown in  Fig.~\ref{fig:hscatxsec}. 

We have also computed the  bound state energies  of the decoupled  massive channel  \eqref{bkapprox}  which occurs in both (\ref{bkchannels}) and (\ref{fvchannels}), 
using the numerical method summarised in the previous section.  They were also computed in  \cite{ForVol:03}.  Our  results are listed in Table \ref{table:fvevalues}, given to 4 significant figures.  They are  in good agreement with the results in  Table II of \cite{ForVol:03}.  Comparing with Fig.~\ref{fig:hscatxsec}, one sees that the energies of the bound states are close to the energy values where the resonances occur. The bound states of the decoupled problem have turned into resonances in the coupled problem. 
This is typical Feshbach behaviour.

\begin{table}[htp]
\centering
\begin{tabular}{cccccccc}
 \hline \hline $n$&1&2&3&4&5&10&15  \\ \hline
 $\omega_n$&0.7984&0.9263&0.9618&0.9766&0.9842&0.9956&0.9980\\
 \hline \hline
\end{tabular}
\label{table:fvevalues}
\caption{Values of $\omega_n$ for the eigenvalue problem   (\ref{bkapprox}).}
\end{table}

Next we turn to scattering states in the system \eqref{bkchannels},  exploiting the equivalence 
with  the  simpler system  \eqref{xizetachannels}.  In the energy range that we are interested in, with $\omega^2<1$,  only 
the equation \eqref{zeta} has scattering solutions. However, from  our discussion after \eqref{susymap} we know that a non-vanishing $\zeta$ violates the background gauge condition. We can therefore conclude that  the system \eqref{bkchannels} has {\em no} scattering states satisfying the background gauge condition in the  range $\omega^2<1$.  This is surprising in view of the superficial similarity  to the system  \eqref{fvchannels},  which, as we saw above,  shows interesting resonance scattering. 

We end this section with  some observations  about  scattering solutions of \eqref{zeta}. Even though they do not satisfy the background gauge condition and therefore do not correspond to solutions of the linearised YMH equations we expect them to play a role in the supersymmetric version of the theory as fermonic scattering states. The scattering problem associated to \eqref{zeta} is also of independent interest since it can be solved exactly. 
  Inserting the expression \eqref{WH} for the profile function $W$, the equation  \eqref{zeta} takes the following simple form:
\begin{equation}
\label{solvablescatt}
 -\frac{d^2\zeta }{dr^2} + \frac{2}{\sinh^2(r)} \zeta  =   \omega^2 \zeta.
\end{equation}
This equation was studied  in the context of  fermion scattering off monopoles in \cite{MarMuz} (the authors there did
not include the Higgs field, but this does enter \eqref{solvablescatt} anyway). A scattering solution regular at the origin is 
\begin{equation}
 \zeta(r) = (i\omega +\coth (r) )e^{-i\omega r} + (i\omega - \coth (r) )e^{i\omega r},
\end{equation}
 from which we can read off the expression 
\begin{equation}
 e^{2i\delta(\omega)}= \frac{i+ \omega }{i-\omega}
\end{equation}
for the phase shift. Hence $
 \sin^2\delta(\omega) = \omega^2/(1+\omega^2)$,
so that the partial scattering cross section is simply
\[
\sigma_0 =   \frac {4\pi}{1+\omega^2}.
\]

\section{Perturbations of the  non-BPS monopole}
\label{nonBPSsect}
We now extend our study of the linearisation around $SU(2)$  monopoles in YMH theory
 by moving away from the BPS  limit and allowing for a non-zero value of $\lambda$ in \eqref{EoM}. 
Outside the BPS limit we cannot use the quaternionic version of the linearised
YMH equations, (\ref{QLYMH}), because it relied on the quaternionic version of the Bogomol'nyi equation, (\ref{QLBE}), which no longer holds. We therefore will not study the most general perturbations in this section but restrict attention to the hedgehog ansatz \eqref{hedgehog} and study perturbations within this ansatz.
With $\lambda\neq 0$ we do not have the
analytic solutions (\ref{WH}) for $W(r)$ and $H(r)$ of the equations  (\ref{2ndOrderWHeqnsH}) for the profile functions in the Hedgehog, so we must solve these numerically. We use the NAG routine
\nag{D02GAF}, which solves non-linear boundary value problems using a finite difference technique with
deferred correction.  Plots for a range of different values of $\lambda$ can be found in \cite{ManSut:04}.
 
As we switch on $\lambda$,  the asymptotic  behaviour of the monopole changes, and this will be crucial in studying perturbations. The asymptotic function   for $W(r)$ is still as in
(\ref{WHasympt}), but $H(r)$  no longer has a constant term as $r\to\infty$. The new
leading terms are 
\begin{equation}\label{WHmassasympt}
W(r)\approx  \frac 1 2 e^{-r} , \qquad 
H(r)\approx  -r + O(e^{-r}) \quad \text{as} \;\;  r\to\infty.
\end{equation}
Even for small values of $\lambda$, integration beyond around $r=10$ becomes unstable, so for use in the linearised equations we spline with the asymptotic approximations for $W(r)$ and $H(r)$.   In studying the linearised problem we fix a value of $\lambda$ and use the profile functions for that case. Thus, from now on we set $\lambda =0.1$.

\begin{figure}[htp]
\centering
\includegraphics[width=10truecm]{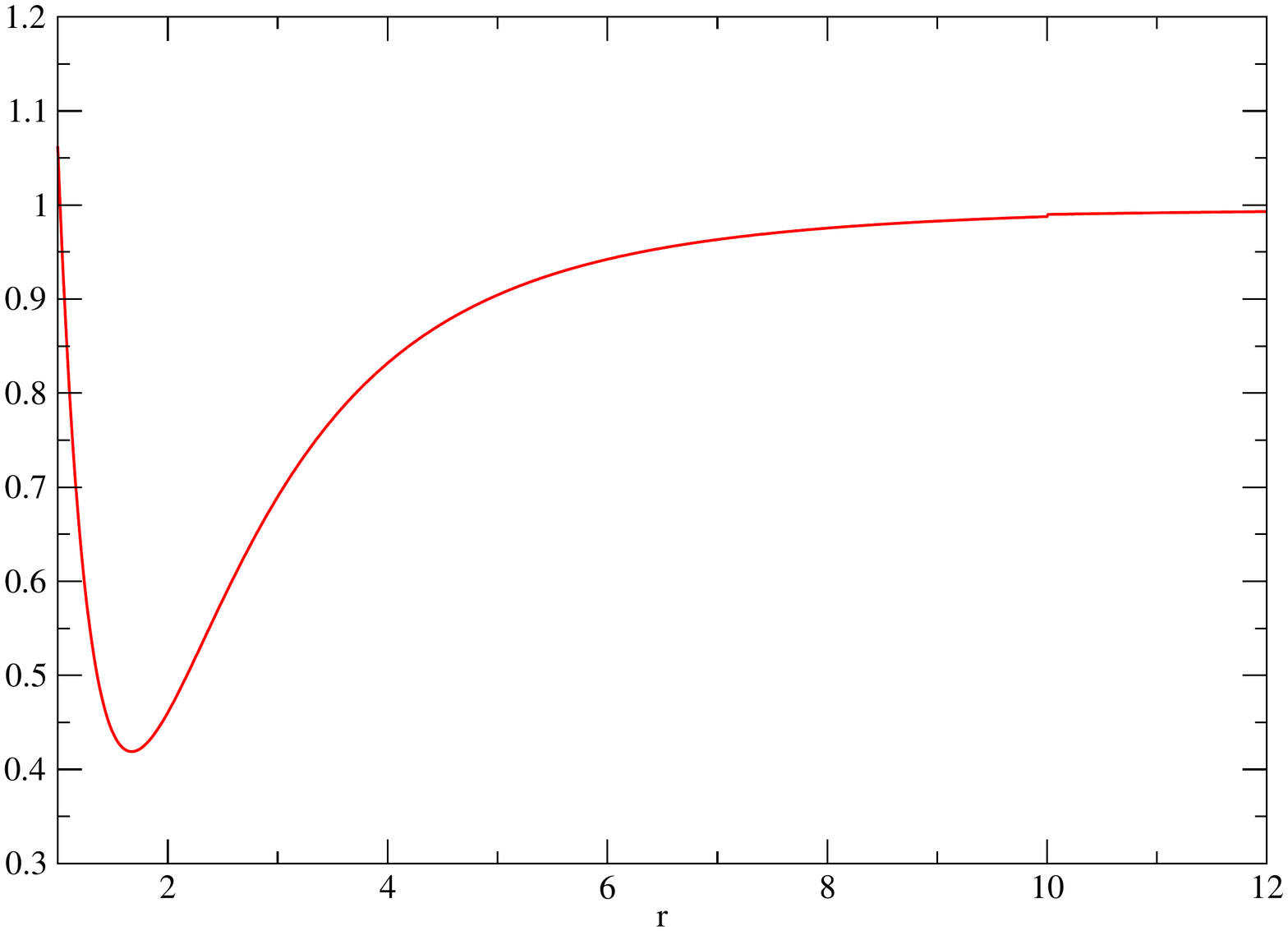}
\caption{The potential $\frac{3W^2+H^2-1}{r^2}$ plotted against $r$, for $\lambda=0.1$. }
\label{fig:massivepot}
\end{figure}

The linearisation of  (\ref{2ndOrderWHeqnsH}) was already given in 
\cite{ForVol:03}. Denoting the hedgehog profile functions by $H^s$ and $W^s$, inserting 
\begin{equation}
W(r,t)=W^s(r)+e^{i\omega t} w(r), \qquad H(r,t)=H^s+\sqrt{2} e^{i\omega t} h(r)
\end{equation}
and keeping only linear terms in $w$ and $h$, one obtains, after re-naming $H^s\to H$, $W^s\to W$, 
\begin{subequations}\label{LWH}
\begin{align}                                                                                                                                              
-\frac{d^2 w}{dr^2} +\frac{3W^2+H^2-1}{r^2}w -2\sqrt{2}\frac{WH}{r^2}h &=\omega^2 w, \label{LWHw}\\
-\frac{d^2h}{dr^2} +\frac{2W^2}{r^2}h  + \lambda \left(\frac{H^2}{r^2}-1\right)h -2\sqrt{2}\frac{WH}{r^2}w &=\omega^2 h. \label{LWHh}
\end{align}
\end{subequations}
We note that (\ref{LWHw}) is the same equation as (\ref{fvw}), but the
differing asymptotic behaviour (\ref{WHmassasympt}) of $H(r)$ means that the potential appearing in this equation
now behaves as 
\begin{equation}
\label{newpotasy}
\frac{3W^2+H^2-1}{r^2} \approx  1-\frac{1}{r^2} + O(e^{-r}) \quad  \text{for large} \; r.
\end{equation} 
This potential is plotted in Fig. \ref{fig:massivepot}. 

\vspace{1cm}

The system \eqref{LWH} has the same generic form as the systems \eqref{fvchannels} and  \eqref{bkchannels}
studied in the previous section.  We follow the same strategy in analysing it as we did  in Sect.~\ref{scatsec}. 
Thus we first  consider the bound state problem, where the right hand side
of (\ref{LWHw}) is set to zero, so that it is artificially decoupled from the
system. The eigenvalue problem we  need to solve is 
\begin{equation}
\left(-\frac{d^2}{dr^2}+\frac{3W^2+H^2-1}{r^2}\right)w=\omega^2w. \label{LWHwdec}
\end{equation}
As we saw in \eqref{newpotasy}, the potential appearing here tends
to the limiting value $1$  faster than in the Coulomb case but still according to an inverse square law,  which is
the threshold case for supporting infinitely many bound states, as discussed in
in  \cite{LanLif:QM}.  As shown there, the radial Schr\"odinger problem   
\begin{equation}\label{gensquarepot}
-\frac{d^2w}{{dr}^2} + V(r)w=E w,
\end{equation}
where $V(r)\sim\frac{\beta}{r^2}$ for large $r$, has infinitely many bound states with $E<0$ if 
$\beta<-\frac{1}{4}$ (for $0>\beta \geq
-\frac{1}{4}$  only a finite
number of bound states are supported).
Thus we expect infinitely many bound states with $\omega^2<1$ in  \eqref{LWHwdec}, with  all but the lowest
values of  $\omega^2$   lying very  close to $1$. This poses problems for the
NAG routine \nag{D02KEF}, which requires a good initial estimate of the eigenvalue.
Fortunately, the inverse square potential is a well-studied problem, and  the WKB method provides an approximation for the ratio of 
successive eigenvalues \cite{MorEltFri:01}. Writing $\omega_n^2$ for the $n$-th eigenvalue in  \eqref{LWHwdec}
 we have the following formula, valid for large $n$:
\begin{equation}\label{moritzformula}
\frac{\omega_n^2-1}{\omega_{n+1}^2-1} = e^{\frac{4\pi}{\sqrt{3}}}.
\end{equation}

\begin{figure}[htp]
\centering
\includegraphics*[width=8truecm]{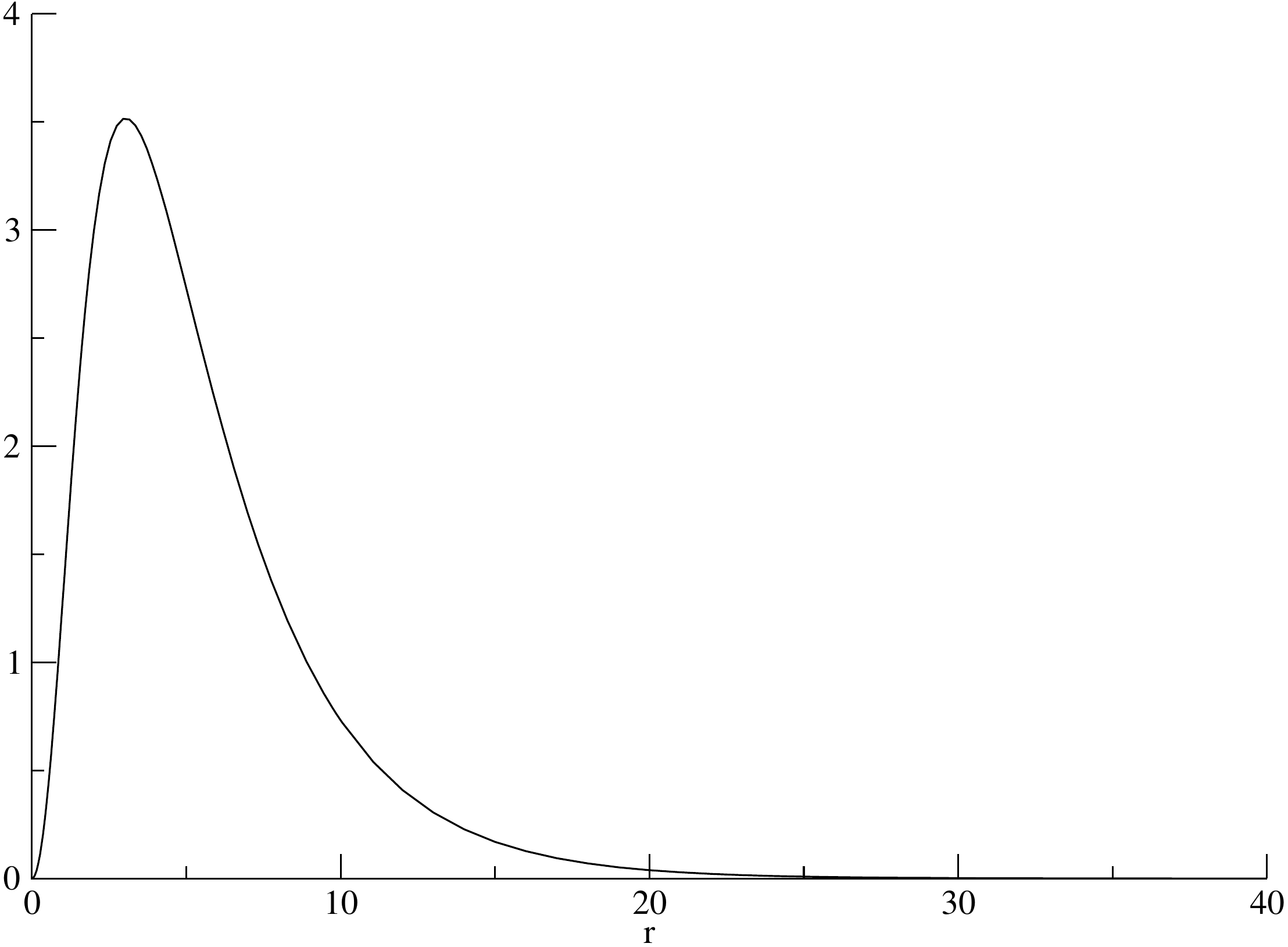} 
\caption{The lowest bound state  of (\ref{LWHwdec}) for $\lambda=0.1$}
\label{fig:k0massive}
\end{figure}

\begin{figure}[htp]
\centering
\includegraphics*[width=6truecm]{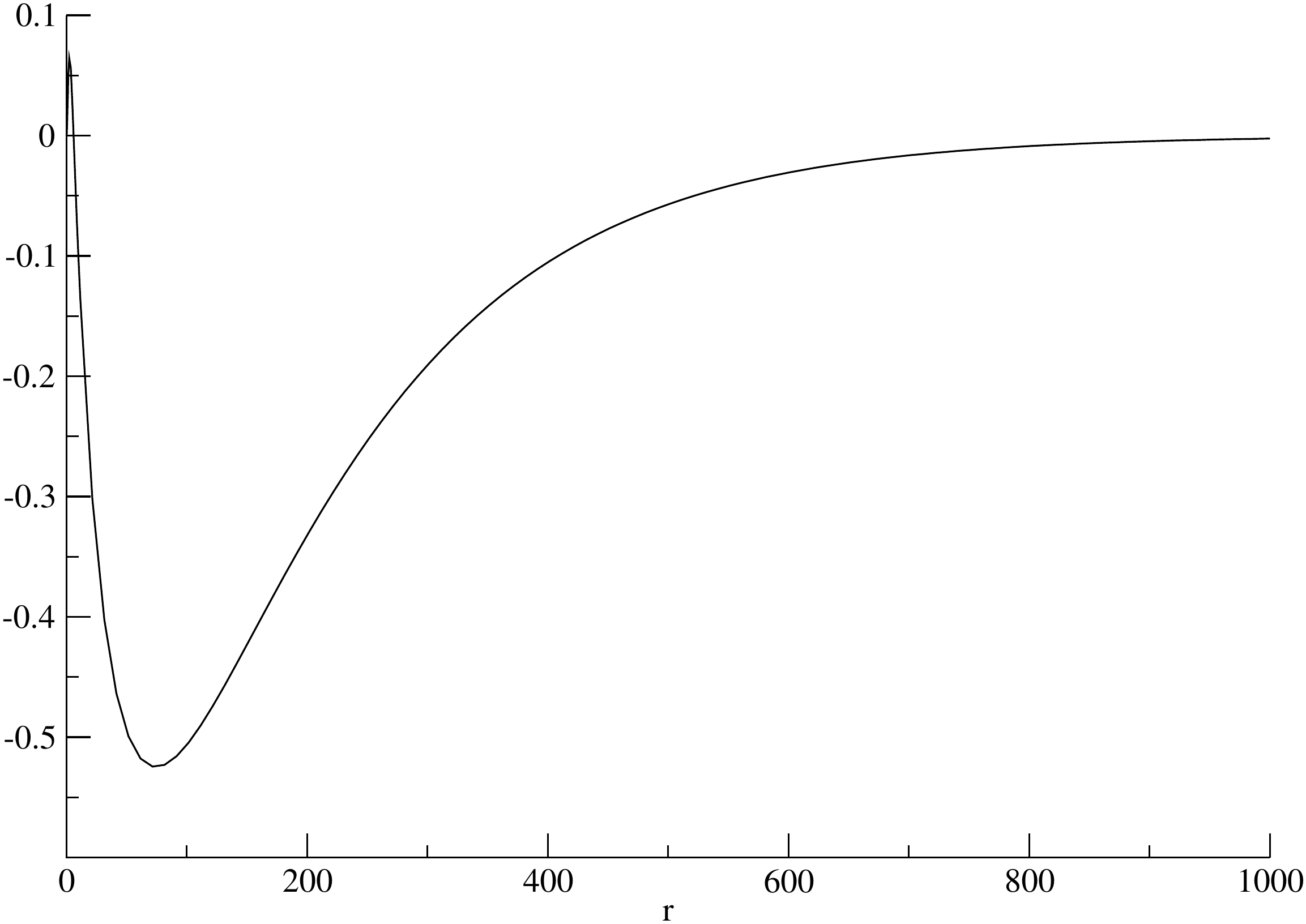}
\includegraphics*[width=6truecm]{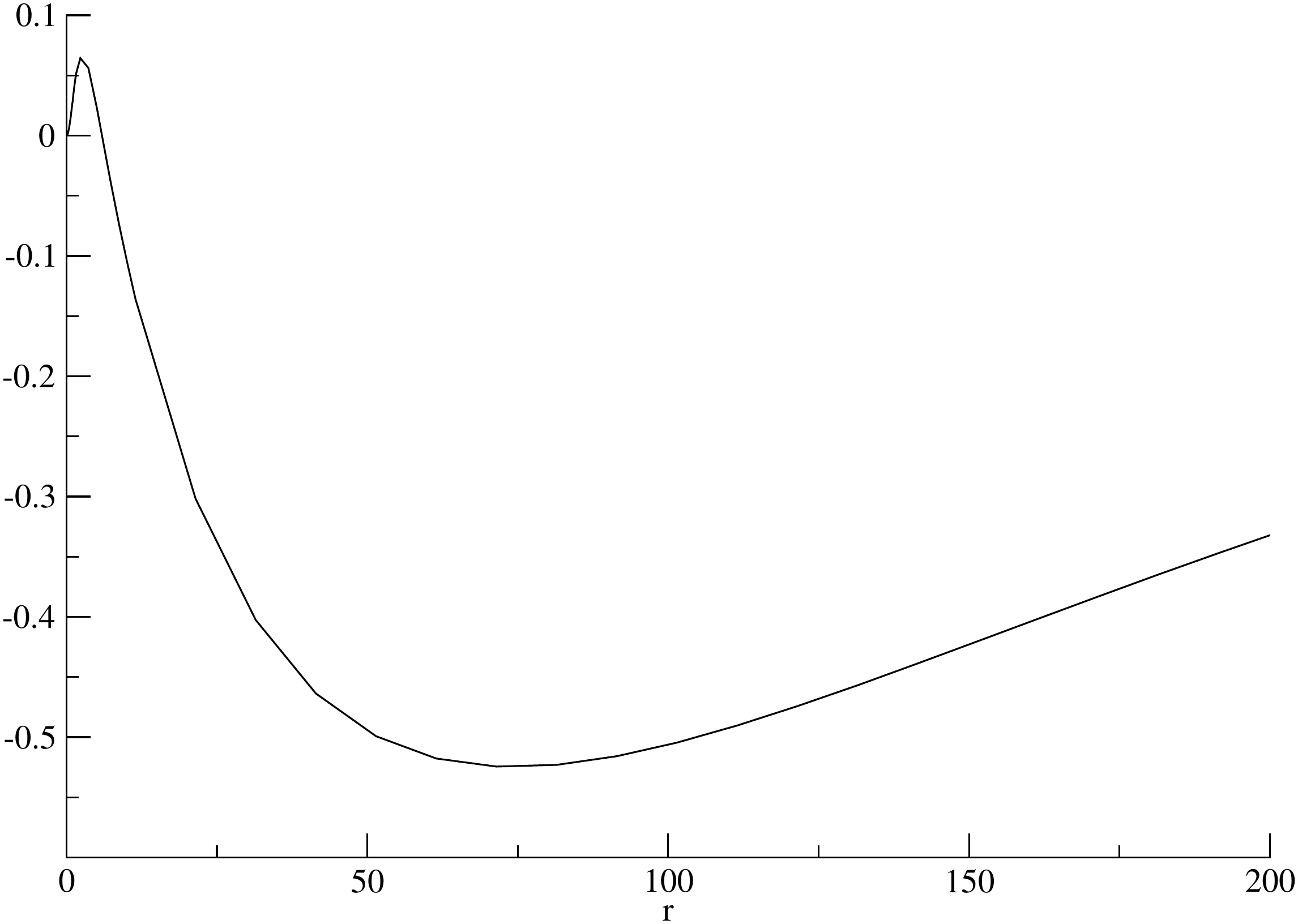} \\
\includegraphics*[width=6truecm]{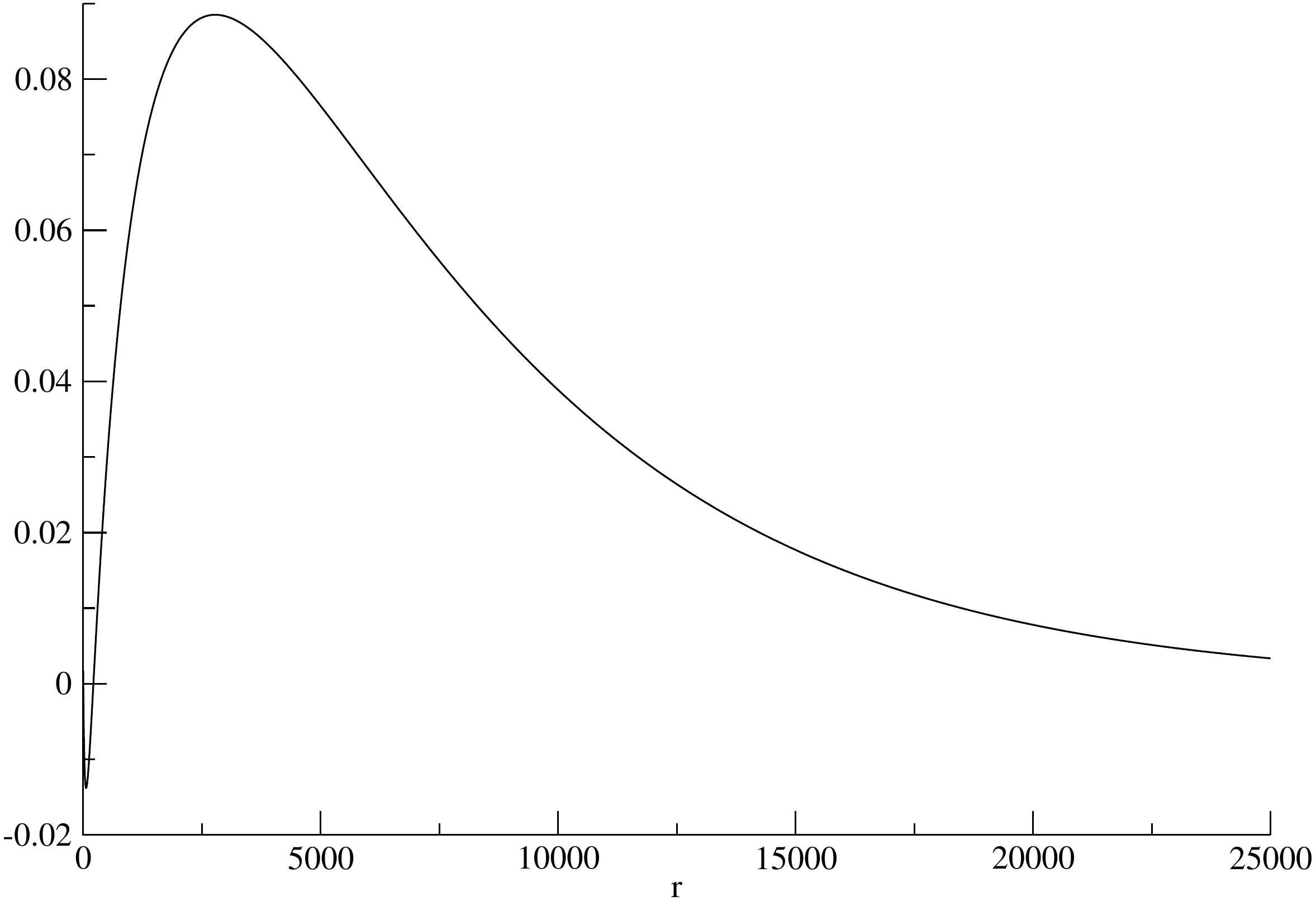}
\includegraphics*[width=6truecm]{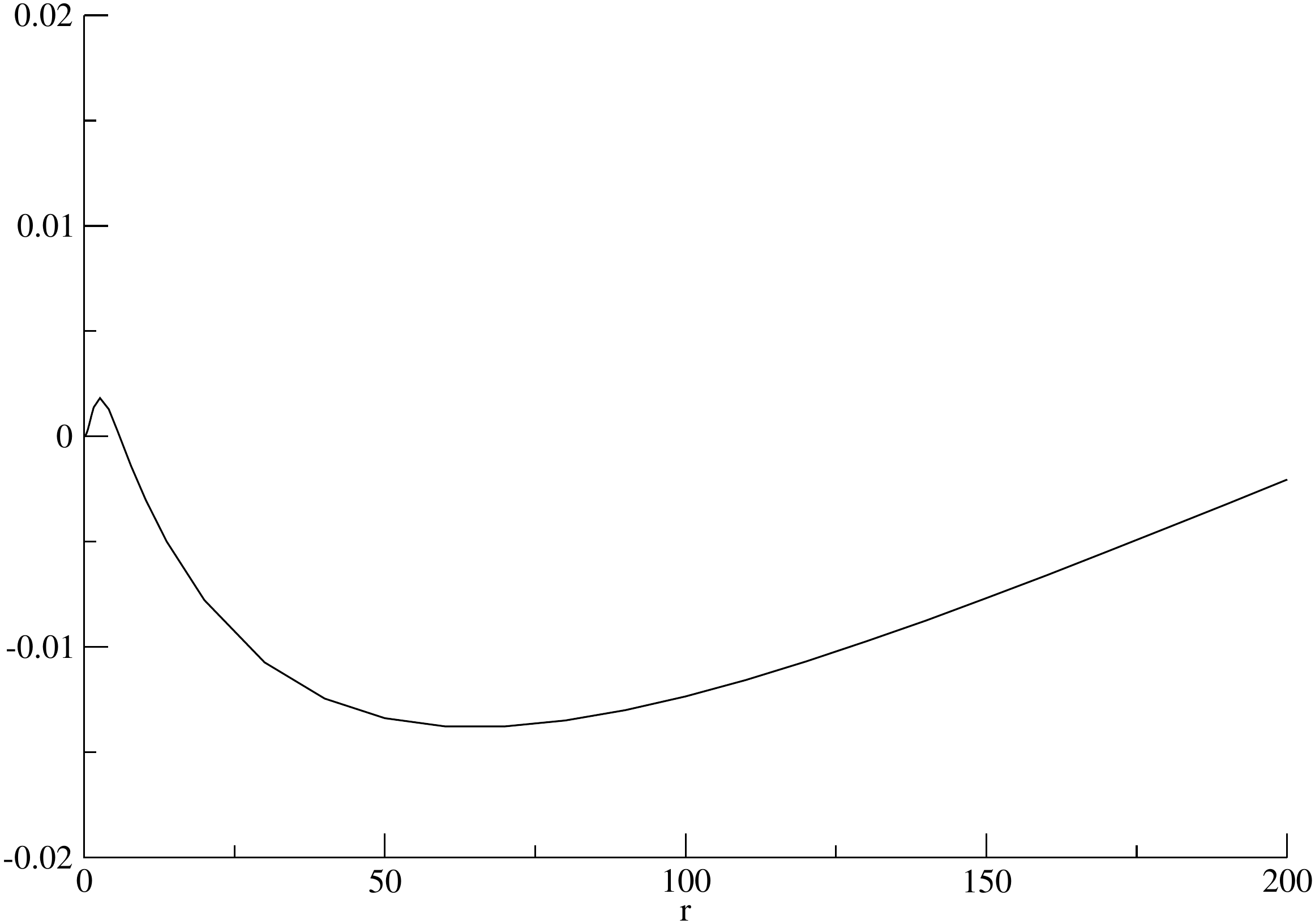}
\caption{The 2nd (top)  and third (bottom)  bound state  of (\ref{LWHwdec}), for $\lambda = 0.1$.
The scale on the right is chosen to show the first oscillation near the origin}
\label{fig:k1k2massive}
\end{figure}

The first eigenvalue and eigenfunction is easily  computed  using the NAG routine \nag{D02KEF}; the eigenfunction is displayed in Fig. \ref{fig:k0massive}. 
Initial guesses for subsequent eigenvalues can then be computed using (\ref{moritzformula}). This
approximation  expected to be good only  for large $n$, but worth using as an initial estimate in 
\nag{D02KEF} for $n=2,3,4$.
The computed eigenvalues, predicted values, and estimated error in computed
value are displayed in Table \ref{table:massiveevalues}.  Rather surprisingly, the WKB guess   even for   $n=3$   is already accurate to 10 decimal
places.
The associated eigenfunctions are plotted in Fig. \ref{fig:k1k2massive}. They  only show  the  exponential decay  characteristic of bound state wavefunctions
 for rather large values of $r$, and we therefore plot them twice, using different scales.  

\begin{table}[htp]
\centering
\begin{tabular}{ccccc}
 \hline \hline $n$&1&2&3&4  \\ \hline
$\omega_n$&0.95343&0.99998&0.999999985&0.999999999990\\
WKB  prediction&n/a&0.99997&0.999999985    &0.999999999989\\
$d\omega_n$&$0.38\times10^{-6}$&$0.68\times10^{-6}$&$0.63\times10^{-10}$&$0.42\times10^{-13}$\\
 \hline \hline
\end{tabular}
\label{table:massiveevalues}
\caption{Eigenvalues $\omega_n$ of (\ref{LWHwdec})  with $\lambda = 0.1$, the predicted values for $n>1$ from
  \ref{moritzformula}, and the estimated accuracy $d\omega_n$ of the computed value $\omega_n$.}
\end{table}

As discussed in appendix  \ref{QNMTheory}, the presence of bound states in \eqref{LWHwdec}
suggests the presence of 
(Feshbach) resonances in the radiative channel of the two channel problem \eqref{LWH}.
We now  look for such resonances by studying  the scattering associated with \eqref{LWH}, in the 
energy range $\omega^2<1$.  We follow the  procedure summarised and used in Sect.~\ref{scatsec}. 
Thus we integrate (\ref{LWH}) for $\lambda=0.1$ and a range of $\omega^2<1$,  using the \nag{D02L} suite of
NAG routines for the integration. These  solve  initial value problems for ODEs   using the Runge-Kutta-Nystrom method. We then tune the initial
conditions with a shooting method to ensure that the
function $w(r)$ decays at large $r$. 
The system has a regular singular point at the origin, which means that the
numerical integration must start a little away from $r=0$. We make a series
expansion of (\ref{LWH})  about $r=0$ to
find the correct initial data. We find that $w(r) \approx A r^2$ and $h(r) \approx B r^2$ for real constants $A,B$, as in the $\lambda=0$ case.

\begin{figure}[htp]
\centering
\includegraphics*[width=7cm]{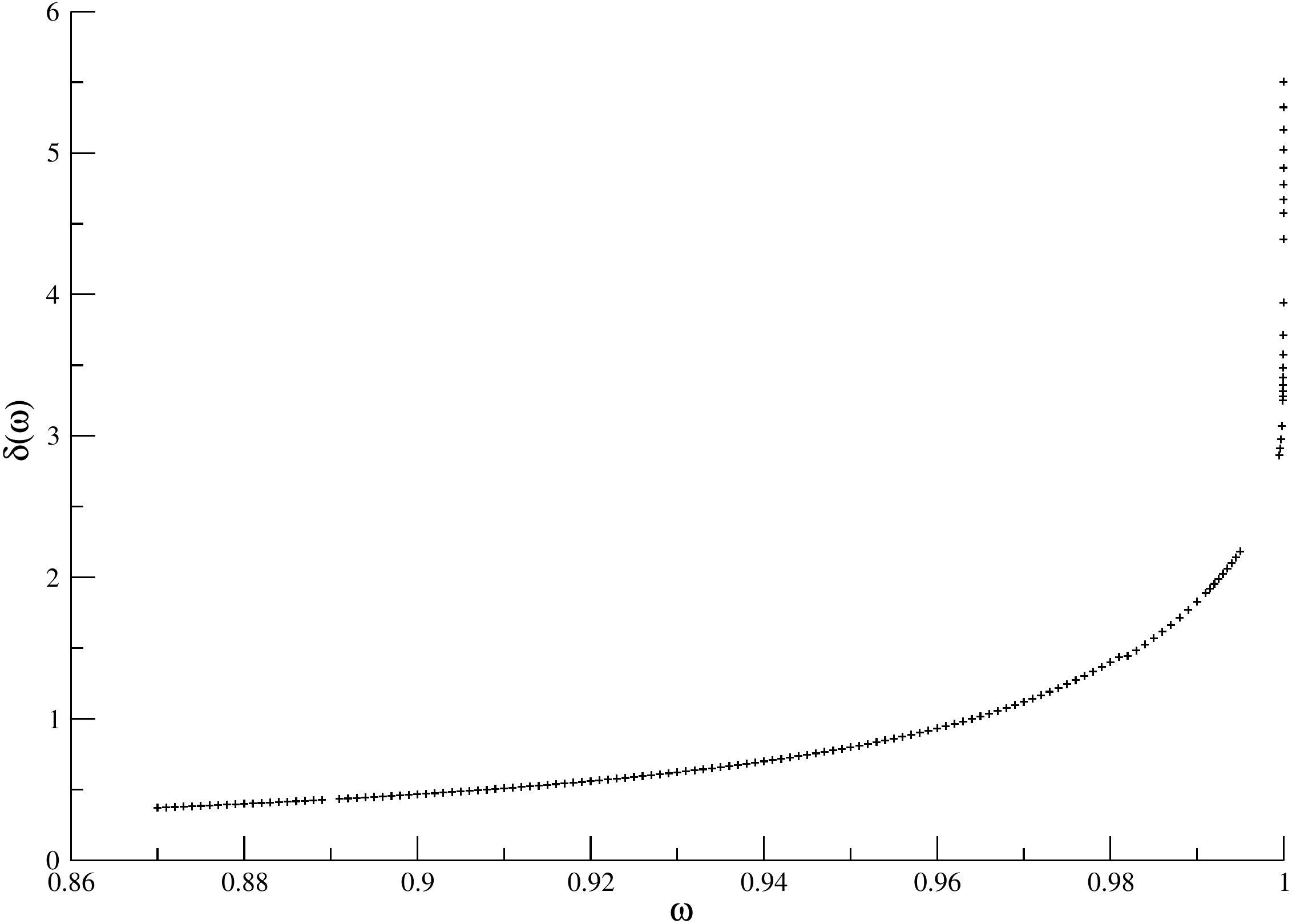}
\caption{ The phase shift $\delta(\omega)$ for the system (\ref{LWH}) }
\label{fig:massivemplephase}   
\end{figure}

We determine the ratio $A/B$ which allows $w$ to decay at infinity and   integrate to compare the solution for $h$ at
large distances to the spherical  Bessel  function  $j_0$,  as in (\ref{mpbesselcomparison}) and to extract the phase shift  $\delta(\omega)$. 
Our results are  shown 
in Fig. \ref{fig:massivemplephase}.
We see that as $\omega\to 1$, the phase
shift $\delta$ increases sharply in value. An increase in $\delta$ by $\pi$ 
suggests a resonance. However,   successive and closely spaced  increases cannot easily  be separated.

\begin{figure}[htp]
\centering
\includegraphics*[width=6cm]{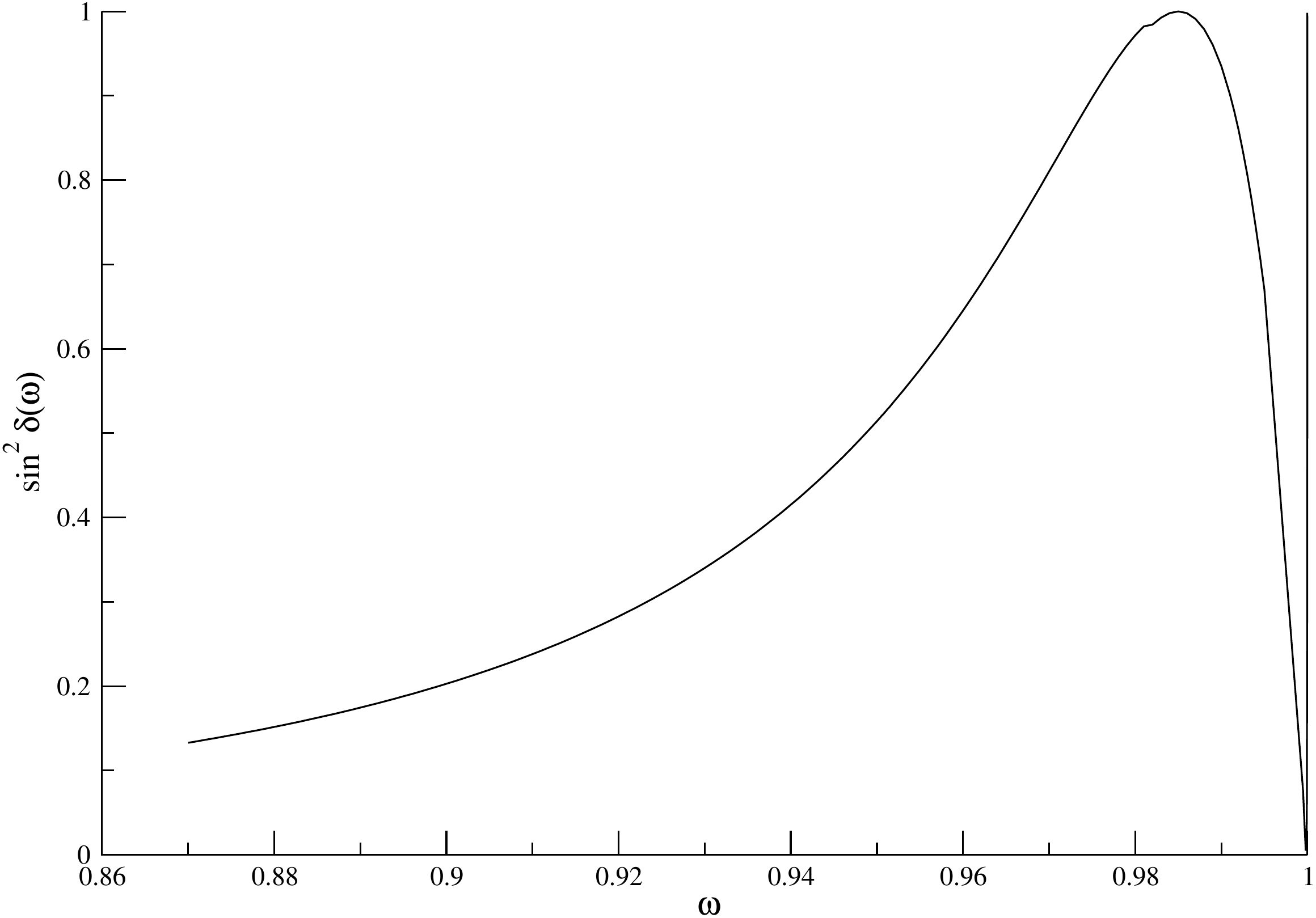}
\includegraphics*[width=6cm]{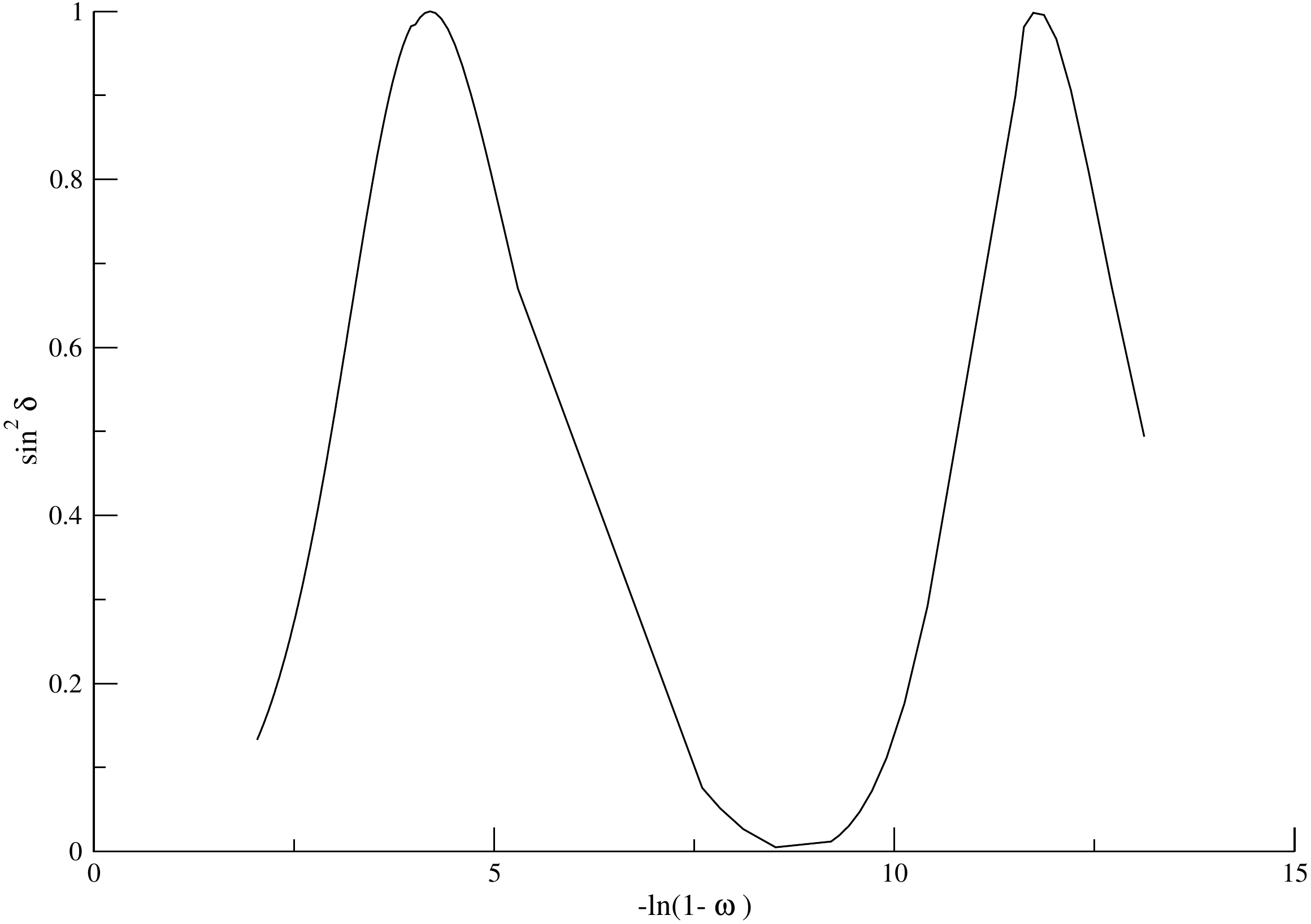}
\caption{For the system (\ref{LWH}), the function $\sin^2\delta$ is plotted    as functions of  $\omega$
(left) and  as a function of  $-\ln(1-\omega)$ (right).}
\label{fig:massivemplexsectlog}   
\end{figure}

In  Fig.
\ref{fig:massivemplexsectlog}  we plot $\sin^2\delta$ as a function of $\omega$. We can clearly see one fairly wide resonance
centred around $\omega=0.985$. Again, the accumulation of the bound state
energies at 1, 
as shown in Table \ref{table:massiveevalues}, makes it difficult to discern
the interesting behaviour as  $\omega\to 1$. 
We stretch out the region near threshold to see if there are, as we expect,
more resonances as $\omega\to 1$. 
The result is plotted on the right in Fig. \ref{fig:massivemplexsectlog}. We can see that
there is at least one more resonant peak. 
We solved (\ref{LWH}) for energies up to $\omega=0.999998$, which takes us
past only the first 
two bound state energies in Table \ref{table:massiveevalues}, thus the
occurrence of two distinct resonance peaks in the scattering cross section is
exactly what is expected.
At higher energies the wavefunctions are extremely long range, of the order of
$r=10^6$. Therefore,  integrating for
higher energies becomes increasingly time consuming. We  expect, however, that 
each of infinitely many bound states of \eqref{LWHwdec} will turn into resonances, and that 
infinitely many more
resonant peaks in the scattering cross section  arise as $\omega\to 1$.

\section{Discussion and conclusion}

The linearised YMH  equations  in the background of a 't Hooft-Polyakov monopole have very interesting spectral properties. We have seen that there 
is a zero-energy bound state and  an  infinity of Coulomb bound states  in the same energy region as  an infinity of  Feshbach resonances.  For vanishing total angular momentum, there are two coupled systems which, despite their superficial similarity, have very different spectral properties: in one (called the Higgs system here) there are infinitely many Feshbach resonances, while in the other (called the photon system here)  there  are infinitely many true bound states,  but no scattering states satisfying the background gauge condition and, in particular, no resonances. The occurrence of bound states can be understood in terms of a decoupling transformation.  We also showed that the Feshbach resonances in the Higgs system persist when the Higgs self-coupling $\lambda$  is switched on. We saw that they   become even more densely spaced, essentially because an attractive $1/r$ potential in the BPS limit is replaced by an attractive $1/r^2$ potential when $\lambda\neq 0$.

In this paper we restricted attention to the  frequency range $\omega^2 <1$  and to spherically symmetric perturbations (in the generalised sense). In this setting,   the scattering is  effectively single channel scattering,
fully described by a single phase shift.  When one goes beyond the threshold $\omega^2 =1 $,  the scattering will be genuine two-channel scattering,
 whose study is more involved, both numerically and in terms of the interpretation of the results. However, very similar multi-channel  scattering 
problems are much studied in atomic and nuclear physics, and  were considered in \cite{Sch:91}  in  the context of monopole scattering, so there is no problem in principle of carrying out a similar study here.  

It would  also be interesting to explore systems arising for larger eigenvalues of the generalised total angular momentum operator $\mathbf{J}$. The combination of our quaternionic formalism with the techniques developed in \cite{JacReb:75,BaiTro:81, Majid} should  provide   an efficient method for finding the corresponding systems of coupled differential equations. As explained in Sect.~\ref{partialsect}, we expect the system for $j=1$ to consist of ten equations, and for $j>2$ of twelve. However, a parity argument will split each of these system into two sub-systems (as it did in our $j=0$ case), so that the largest system one needs to consider has six channels.

To end, we point out  the striking similarity between the spectral properties of the linearised YMH system in the background of the 't Hooft-Polyakov monopole and those of the Laplace operator on the moduli space of charge two $SU(2)$ monopoles \cite{Sch:91,ManSch:92}. The latter also include zero-energy bound states, Coulomb bound states embedded in the continuum and Feshbach resonance scattering.  These similarities are likely to have an interpretation in terms
 of electric-magnetic duality. Spelling this out is left as the topic for a future investigation.

\appendix

\section{Feshbach resonances}\label{QNMTheory}

A Feshbach resonance \cite{Feshbach} is a resonance  in a system consisting of several channels
in which a bound state occurs if the coupling(s) between the channels vanishes.  Feshbach resonances 
are much studied in the context  atomic physics \cite{Mit:66}  but also occur in other contexts,
see \cite{Milstein} for a pedagogical and recent account. 

Consider a simple system consisting of two channels with coupling terms including, for convenience, a parameter $0\leq q \leq 1$,
\begin{align}\label{coupledgen}
   -\frac 1 r \frac{d^2}{dr^2} (r u) + V(r)u + qC(r)v& =E u, \nonumber \\
   -\frac 1 r \frac{d^2}{dr^2} (r v)  +\hat{V}(r)v + qC(r)u&=E  v.
 \end{align}
We can decouple the equations by setting the parameter $q$ to zero. With $V<0$,
suppose that the
eigenvalue problem
\begin{equation}\label{decoupledgen}
- \frac 1 r \frac{d^2}{dr^2} (r u) + V(r)u=E  u
\end{equation} 
has bound states for $E< 0 $, occurring at
$E_0, E_1, E_2, \ldots$ but that those values  are part of the continuous spectrum for  the other decoupled equation
\begin{equation}\label{decoupledgenn}
- \frac 1 r \frac{d^2}{dr^2} (r v) + \hat{V}(r)v=E  v.
\end{equation} 
When the coupling parameter $q$ is non-zero, the  bound states in \eqref{decoupledgen} can leak into the radiative channel \eqref{decoupledgenn}
and decay.  When that happens,  Feshbach resonances in 
(\ref{coupledgen}) generically occur at values close to $E_0, E_1, E_2,
\ldots$.  Examples of  such resonances are studied 
in  \cite{ForVol:03} and  \cite{Sch:91}, both in the context of magnetic monopoles. A single channel eigenvalue problem
     such as (\ref{decoupledgen}) is generally much easier to solve
     than a two channel coupled problem such as
     (\ref{coupledgen}). Hence, it is worth looking at the bound state
     problem, as it gives an idea of where the resonances will
     occur. In systems where the coupling terms are weaker at large
     distances than the potential, the decoupled problem is a good
     approximation. Then these preliminary calculations mean that the region of
     searching for resonances is narrowed down, so that 
     computational time is reduced.

\end{document}